\documentclass[journal,twocolumn]{IEEEtran}
\usepackage{multirow}
\usepackage{booktabs}
\usepackage{amsfonts}
\usepackage{placeins}
\usepackage{array}
\usepackage{amsmath,cases,amssymb}
\usepackage{amsmath}
\usepackage{mathtools}
\usepackage{graphicx}
\usepackage{subcaption}
\usepackage{cite}
\usepackage{epsfig}
\usepackage{todonotes}
\usepackage{url}
\usepackage{algorithm}
\usepackage{algorithmic}
\usepackage{epstopdf}
\usepackage{balance}
\usepackage{color}
\usepackage{algorithm}
\DeclareGraphicsExtensions{.eps,.pdf}

\usepackage{scalerel}
\usepackage{multicol}
\usepackage{amssymb}% http://ctan.org/pkg/amssymb
\usepackage{pifont}% http://ctan.org/pkg/pifont
\usepackage{cases}

\usepackage{scalerel}
\usepackage{orcidlink}

\let\labelindent\relax
\usepackage{enumitem}

\usepackage{mathbbol}
\usepackage{cite}
\usepackage[T1]{fontenc} % optional
\usepackage{amsthm} % format theorem (bold title, italic content)
\usepackage{eqparbox}
\usepackage{textcomp}
\usepackage[eulergreek]{sansmath}
\usepackage{color}
\usepackage[cmintegrals]{newtxmath}

\usepackage[open,openlevel=1]{bookmark}
\usepackage{ctable}
\usepackage{threeparttablex}

\newtheorem{remark}{Remark}
\newtheorem{proposition}{Proposition}

\allowdisplaybreaks

\newcommand{\bseq}{\begin{subequations}}
\newcommand{\eseq}{\end{subequations}}
\newcommand{\baln}{\begin{align}}
\newcommand{\ealn}{\end{align}}
\newcommand{\balnd}{\begin{aligned}}
\newcommand{\ealnd}{\end{aligned}}
\newcommand{\beq}{\begin{equation}}
\newcommand{\eeq}{\end{equation}}
\newcommand{\beqn}{\begin{eqnarray}}
\newcommand{\eeqn}{\end{eqnarray}}
\newcommand{\beqno}{\begin{eqnarray*}}
\newcommand{\eeqno}{\end{eqnarray*}}
\newcommand{\bma}{\begin{displaymath}}
\newcommand{\ema}{\end{displaymath}}
\newcommand{\bnu}{\begin{enumerate}}
\newcommand{\enu}{\end{enumerate}}
\newcommand{\bce}{\begin{center}}
\newcommand{\ece}{\end{center}}
\newcommand{\btb}{\begin{tabular}}
\newcommand{\etb}{\end{tabular}}
\newcommand{\bieq}{\begin{IEEEeqnarray}}
\newcommand{\eieq}{\end{IEEEeqnarray}}

\newcommand{\st}{{\mathrm{s.t.}}}
\newcommand{\subnum}{\IEEEyessubnumber}

\makeatletter
\newcommand{\linebreakand}{%
\end{@IEEEauthorhalign}
\hfill\mbox{}\par
\mbox{}\hfill\begin{@IEEEauthorhalign}
}
\makeatother

\begin{document}
\title{%Joint Sum-rate and Handover Optimization for C-Band Integrated Satellite and Terrestrial Networks Supporting Urban Seamless 5G Automotive Connectivity
% Enhanced Throughput and Seamless Handover Solutions for Urban 5G-Automotive Supported C-Band Integrated Satellite-Terrestrial Networks
Enhanced Throughput and Seamless Handover Solutions for Urban 5G-Vehicle C-Band Integrated Satellite-Terrestrial Networks }
% \vspace{-3mm}

\author{ \IEEEauthorblockN{Hung Nguyen-Kha\orcidlink{0000-0002-5956-4279}, \IEEEmembership{Graduate Student Member,~IEEE,} Vu Nguyen Ha\orcidlink{0000-0003-1325-3480}, \IEEEmembership{Senior Member,~IEEE,}\\ Eva Lagunas\orcidlink{0000-0002-9936-7245}, \IEEEmembership{Senior Member,~IEEE,} Symeon Chatzinotas\orcidlink{0000-0001-5122-0001}, \IEEEmembership{Fellow,~IEEE},\\ and Joel Grotz\orcidlink{0000-0002-4095-4015}}, \IEEEmembership{Senior Member,~IEEE }

\IEEEcompsocitemizethanks{ 
This work has been supported by the Luxembourg National Research Fund (FNR) under the project INSTRUCT (IPBG19/14016225/INSTRUCT).
% This research was funded in whole, or in part, by the Luxembourg National Research Fund (FNR) through the Project INtegrated Satellite – TeRrestrial Systems for Ubiquitous Beyond 5G CommunicaTions (INSTRUCT) under Grant IPBG19/14016225/INSTRUCT. 
The preliminary result of this manuscript was presented in IEEE VTC Fall'24 \cite{Hung_VTC24}.
}

\IEEEcompsocitemizethanks{H. Nguyen-Kha, V. N. Ha, E. Lagunas, and S. Chatzinotas are with the Interdisciplinary Centre for Security, Reliability and Trust (SnT), University of Luxembourg, 1855 Luxembourg Ville, Luxembourg.  (e-mail: khahung.nguyen@uni.lu; vu-nguyen.ha@uni.lu; eva.lagunas@uni.lu; Symeon.Chatzinotas@uni.lu).}

\IEEEcompsocitemizethanks{J. Grotz is with SES, Chateau de Betzdorf, Betzdorf 6815, Luxembourg (e-mail: Joel.Grotz@ses.com).}
\vspace{-10mm}
}

\markboth{ACCEPTED FOR PUBLICATION IN IEEE TRANSACTIONS ON COMMUNICATIONS}{}

\maketitle

\begin{abstract} 
This paper investigates downlink transmission in 5G Integrated Satellite-Terrestrial Networks (ISTNs) supporting automotive users (UEs) in urban environments, where base stations (BSs) and Low Earth Orbit (LEO) satellites (LSats) cooperate to serve moving UEs over shared C-band frequency carriers.
Urban settings, characterized by dense obstructions, together with UE mobility, and the dynamic movement and coverage of LSats pose significant challenges to user association and resource allocation. 
To address these challenges, we formulate a multi-objective optimization problem designed to improve both throughput and seamless handover (HO). Particularly, the formulated problem balances sum-rate (SR) maximization and connection change (CC) minimization through a weighted trade-off by jointly optimizing power allocation and BS-UE/LSat-UE associations over a given time window.
This is a mixed-integer and non-convex problem which is inherently difficult to solve. To solve this problem efficiently, we propose an iterative algorithm based on the Successive Convex Approximation (SCA) technique. Furthermore, we introduce a practical prediction-based algorithm capable of providing efficient solutions in real-world implementations. Especially, the simulations use a \textit{realistic 3D map of London} and UE routes obtained from the Google Navigator application to ensure practical examination. Thanks to these realistic data, the simulation results can show valuable insights into the link budget assessment in urban areas due to the impact of buildings on transmission links under the blockage, reflection, and diffraction effects. Furthermore, the numerical results demonstrate the effectiveness of our proposed algorithms in terms of SR and the CC-number compared to the greedy and benchmark algorithms.
\end{abstract}
\vspace{-1mm}
\begin{IEEEkeywords}
%\vspace{-.3cm}
LEO Constellation, Integrated Satellite-Terrestrial Networks, Seamless Connectivity, C-Band, 5G Automotive, Resource Allocation. 
\end{IEEEkeywords}

%\vspace{-1cm}

%\newpage

\vspace{-4mm}
\section{Introduction}
\vspace{-2mm}
\IEEEPARstart{I}{n} recent years, wireless communication networks have rapidly evolved to meet increasing traffic demands and connectivity needs. Consequently, terrestrial networks (TNs) have been swiftly deployed to accommodate this growth. However, TNs often struggle to maintain the connectivity and quality of service for a vast number of devices. Therefore, the next-generation network aims to provide extensive and seamless connectivity, addressing the high traffic demands and service requirements of new applications \cite{SurveyTut_Roadto6G, ProIEEE_6GVision_Challenge_Opp, Mag_Toward6G_Usecase_Technol}. 
{Especially, these requirements become more crucial in urban environments that generally have high traffic and user demand.} A straightforward solution is the dense deployment of base stations (BSs) to enhance network coverage and capacity. However, this approach can be challenging and costly to implement.
Additionally, satellite communication (SatCom) emerges as a promising solution to expand network coverage and capacity \cite{VuHa_ICC23}. 
Particularly, due to the ubiquitous and high capacity, SatCom can provide complementary coverage, aid in offloading for TNs, and jointly serve with TNs.
This approach is supported by the United States (US) Federal Communications Commission (FCC), which has proposed complementary coverage for TNs from space segments \cite{FCC2322_Supplemental_Space_Coverage}.
Although satellite coverage is typically seen as the solution for areas lacking TN connectivity, using SatCom in urban environments, rather than relying on dense base station (BS) deployment, presents a cost-effective alternative. This approach not only enhances coverage but also helps traffic offloading, addressing issues of susceptibility to blockage and congestion events, respectively.
While SatCom was previously expensive, costs have now decreased, especially for Low Earth Orbit satellites (LSats). Hence, reasonable manufacturing and launching costs of LSats led to many interests in developing LEO constellations such as Starlink and OneWeb. Compared to geostationary and medium-earth orbit satellites, LSats offer higher channel gain and lower latency due to their lower altitude, making the integration of satellite and TN systems more feasible \cite{Hung_2tierLEO}. 

Traditionally, TNs and SatCom Networks (SatNets) have operated in distinct frequency bands (RFBs) due to spectrum regulation. However, to meet the growing demand for connectivity, both networks have expanded their spectrum utilization, resulting in the development of coexistence radio-access systems. For instance, many countries have extended the TN radio frequency (RF) spectrum into the millimeter-wave frequency and C-band ($3.4-4.2$ GHz). {Concurrently, significant efforts from academia and industry aim to lower the RFBs of Non-Terrestrial Networks (NTNs), including SatNets, from Frequency Range 2 (FR2) to Frequency Range 1 (FR1), incorporating bands such as the S-band and L-band \cite{3gpp.38.101-5}, which can offer better penetration through precipitation and a longer propagation distance compared to higher RFBs}. Following this trend, Mediatek has proposed scenarios in which the TN spectrum is shared with NTNs under complementary management \cite{3gpp.Tdoc_RWS230110}. Notably, various coexistence scenarios between TNs and NTNs have been investigated by 3GPP \cite{3gpp.38.863}. 
% Despite these efforts, spectrum allocation remains underutilized due to varying demand across different geographic areas, with over-utilization occurring only in regions with high user (UE) demand and connectivity \cite{Valenta_CROWNCOM2010}.

In light of the evolving trend toward Integrated Satellite-Terrestrial Networks (ISTNs), sharing the same RFBs between TN and NTN systems is considered a promising solution to enhance spectrum efficiency and increase overall network capacity. 
Based on the agreement between terrestrial and satellite operators, BSs and LSats can work together in ISTNs to provide Internet connectivity. This approach is further supported by the FCC, which has proposed the co-primary use of mobile and satellite services within the same RFBs \cite{FCC2322_Supplemental_Space_Coverage}. 
Furthermore, direct-to-device connectivity has attracted significant attention and is supported by the European Space Agency \cite{ESA_D2D_conf}.
Hence, this approach offers the advantage of improving seamless connectivity by allowing UEs to transit smoothly between TN and NTN systems \cite{3gpp.38.863, FCC2322_Supplemental_Space_Coverage}.
% In light of the evolving trend toward Integrated Satellite-Terrestrial Networks (ISTNs), sharing the same RFBs between TN and NTN systems is considered a promising solution to enhance spectrum efficiency and increase overall network capacity. This approach is further supported by the FCC, which has proposed co-primary use of mobile and satellite services within the same bands \cite{FCC2322_Supplemental_Space_Coverage}. Additionally, it offers the advantage of improving seamless connectivity by enabling UEs to transition smoothly between TN and NTN systems \cite{3gpp.38.863, FCC2322_Supplemental_Space_Coverage}.
Moreover, the C-band has emerged as a critical RFB for future networks. Historically reserved for Fixed Satellite Services (FSS), the C-band is now attracting interest from TN operators due to its wide coverage and high capacity. In the US, the FCC has implemented a plan to repurpose $280$~MHz of C-band spectrum from FSS downlink (DL) to 5G, requiring satellite companies to vacate this portion by December $2023$ \cite{FCC20_Cband_5G}. In Europe, the Radio Spectrum Policy Group has identified three key bands for 5G deployment: $700$~MHz for broad coverage, $3.4$~GHz for a balance between coverage and capacity, and $26$~GHz for high-speed data rates, with the $3.4$~GHz C-band designated as the primary band for implementation. Besides, 3GPP defined specific C-band frequencies ($n77$, $n78$, $n79$) for 5G-NR \cite{3gpp.38.104}, underscoring the C-band's pivotal role in the future of ISTNs.

Although LSats in ISTNs offer substantial benefits, such as enhanced capacity, improved coverage, and seamless connectivity, their inherent movement introduces significant dynamic challenges. 
The extensive coverage provided by satellites also introduces heterogeneity in ISTNs in terms of propagation environments due to the diverse characteristics of the TN areas (urban, suburban, rural) and even environmental differences within a single urban area \cite{Hung_MeditCom24}. This environmental diversity necessitates accounting for the varying impact of the environment within the ISTN's coverage in system design. 
Additionally, UE mobility, particularly the movement of vehicles in urban areas, further complicates the network dynamics \cite{Hung_VTC24}. Although utilizing the same RFB in ISTNs, such as the C-band for DL as per the current spectrum regulation, can enhance spectrum efficiency, increase system capacity, and support seamless connectivity, it also introduces critical cross-interference issues between TN and NTN systems. Consequently, system design must carefully address seamless handovers, cross-system interference, and resource allocation, factoring in the unique characteristics of urban environments and realistic UE mobility to ensure these challenges are adequately reflected in practical deployments.

\vspace{-3mm}
\subsection{Related Works}
\vspace{-1mm}
Recently, the analysis of interference in TN-NTN coexistence and spectrum sharing systems has been investigated in many works \cite{3gpp.38.863, Eva_Access20, Jumaily_TVT22, okati_arXiv2401.08453_Sband, kim_spectrum_2024}. 
In \cite{3gpp.38.863}, the 3GPP has considered various TN-NTN coexistence scenarios wherein TN and NTN systems operate in adjacent bands in the S-band. Subsequently, the throughput loss due to adjacent channel interference was investigated.
In \cite{Eva_Access20, Jumaily_TVT22}, the coexistence systems in C-band were studied. Particularly, the authors in \cite{Eva_Access20} analyzed the impact of TNs on DL FSS systems wherein the aggregated interference of 5G BSs at the ground FSS receiver was investigated. Subsequently, the switch-off and power-back-off schemes are proposed to reduce the interference from critical BSs.
The authors in \cite{Jumaily_TVT22} conducted the measurement of the interference power caused by TN BSs to the FSS receiver. Subsequently, the impact of various experimental deployment scenarios of the tuner and filter on the interference has been carried out. 
Additionally, in \cite{okati_arXiv2401.08453_Sband, kim_spectrum_2024}, the performance analysis is studied based on the stochastic geometry.
In particular, the authors in \cite{okati_arXiv2401.08453_Sband} studied different coexistence scenarios in the S-band, where the system performances in terms of coverage and rate probability were analyzed.
In \cite{kim_spectrum_2024}, the authors studied a system-level performance analysis for the sharing systems, where the ergodic spectrum efficiency of LEO-SatNets is provided when the SatNet spectrum is shared with TNs.
However, in these works, the impact of complex environments, such as in urban areas, is not considered. Furthermore, these works focus on snap-shot model analysis without considering the dynamics of systems.

Besides, the resource allocation in spectrum-sharing systems has been studied in many works \cite{du_JSAC2018, zhang_JSAC2022, Lee_TVT23, Li_TWC2024, martikainen_WoWMoM2023, Zhu_TWC24_BeamManage}.
In \cite{du_JSAC2018, zhang_JSAC2022, Lee_TVT23, Li_TWC2024}, the snap-shot-based spectrum-sharing scenarios for ISTNs have been studied.
In particular, the authors in \cite{du_JSAC2018} studied cooperative and competitive scenarios between TN and SatNet systems, where the second-price auction mechanism is designed to achieve equilibrium between channel sharing from TNs and offloading ability from SatNets.
In \cite{zhang_JSAC2022}, the authors studied the NOMA ISTNs where one satellite provides the backhaul link for BSs and the access link for UEs in the C-band. The system bandwidth (BW) is shared between the backhaul and access links. Subsequently, the user association (UA), BW allocation, and power allocation (PA) are optimized under the QoS and backhaul link constraints.
In \cite{Lee_TVT23}, the author examined various TN-NTN spectrum-sharing scenarios. Subsequently, the interference-aware and group-sharing mechanisms are studied for NTN UL transmission in the reverse pairing spectrum sharing scenario, which aims to group and assign BSs and UEs to resource blocks.
The authors in \cite{Li_TWC2024} studied the rate splitting multiple access (RSMA)-based spectrum sharing systems, wherein the SR maximization problems in RSMA systems are considered for underlay and overlay spectrum sharing scenarios. However, the snapshot-based models in these works can not consider the dynamics of networks, especially the LEO-SatNets. Furthermore, these studies are limited to single-satellite scenarios.

Subsequently, the time-varying systems have been studied in \cite{martikainen_WoWMoM2023, Zhu_TWC24_BeamManage}.
In \cite{martikainen_WoWMoM2023}, authors proposed a dynamic spectrum-sharing mechanism in which bandwidth allocation for each network is determined based on its load, with coordination managed by a spectrum management server.
The authors in \cite{Zhu_TWC24_BeamManage} investigated beam management and scheduling in coexistence networks, focusing on LSat transmissions to satellite cells. In this scenario, TNs and SatNets operate in separate RFBs, with the TN band temporarily shared with SatNets. To address this, the study proposed sequential solutions for handover, beam selection, and spectrum sharing. However, the analysis was limited to the system level, without considering BS/LSat-to-UE transmissions, which may pose challenges in accounting for UE mobility.
Furthermore, existing studies on coexistence and spectrum-sharing systems predominantly rely on statistical channel models, which limit their ability to accurately capture environmental characteristics in the system model.
Although the TN-based UA/PA problems have been extensively studied in literature such as \cite{AbantoLeon_TWC22, Tam_TWC17}, the joint optimization of BS/LSat-UA/PA in co-primary spectrum-sharing ISTNs, incorporating a realistic channel model based on an actual three-dimensional (3D) urban map, remains largely unexplored. This paper aims to address this research gap.

\vspace{-2mm}
\subsection{Research Contributions}
\vspace{-1mm}
In this work, we study a time-window (TW)-based UA and PA mechanism for ISTNs in urban environments wherein TNs and SatNets use the same RFB in the C-band.
The main contributions of this work can be summarized as follows.
\begin{itemize}
    \item We analyze time-varying systems and transmission models for ISTNs operating in the same RFB. Particularly, the utilized channel and antenna models are especially compatible with the ray-tracing (RT) mechanism and the actual 3D maps, which can apply to various areas thanks to the 3D map availability in the \textit{OpenStreetMap} resources \cite{OSM}. Based on which we formulate a multi-objective optimization problem that aims to maximize the sum-rate (SR) while minimizing the number of connection state changes (CCs), with joint optimization of BS/LSat-UA and PA. This problem is categorized as mixed-integer non-convex programming, which presents challenges for direct solutions.
    \item To address this problem, we propose two iterative algorithms using the successive-convex-approximation (SCA) method. The first algorithm can provide the solution over the entire TW but requires knowledge of the channel gain for the entire TW, which can be challenging in practice. For practical implementation, we further introduce the second prediction-based iterative algorithm. Specifically, we develop an RT-based mechanism to sequentially predict the channel gain, enabling the solution of consecutive sub-TW problems.
    \item For practical simulations, we utilize a realistic 3D map of London to assess the impact of the urban environment on ISTN channels and performance. Additionally, UE mobility and network load are modeled based on actual routes from Google Navigator and population data. The simulation results provide valuable insights into the impact of the urban environment with the blockage, reflection, and diffraction effects on the time-varying link budget. Moreover, the numerical results demonstrate the superiority of our proposed algorithms over benchmark schemes in solving the joint SR-CC optimization problem. Furthermore, the small gap in performance between the two proposed algorithms highlights the effectiveness of the prediction-based approach.
\end{itemize}
The preliminary result of this work were presented in \cite{Hung_VTC24} where only the UA for SR maximization problem was studied. The rest of this paper is organized as follows. Section~\ref{sec: sys model} presents the system model and the problem formulation. Section~\ref{sec: propose sol} and \ref{sec: benchmark, complexity} show the proposed algorithms, benchmarks, and the complexity analysis. Section~\ref{sec: results} and \ref{sec: concl} provide the simulation and numerical results and the conclusion, respectively.

\vspace{-4mm}
\section{System Model} \label{sec: sys model}
\vspace{-1mm}
\begin{figure}
    \centering
    \includegraphics[width=85mm]{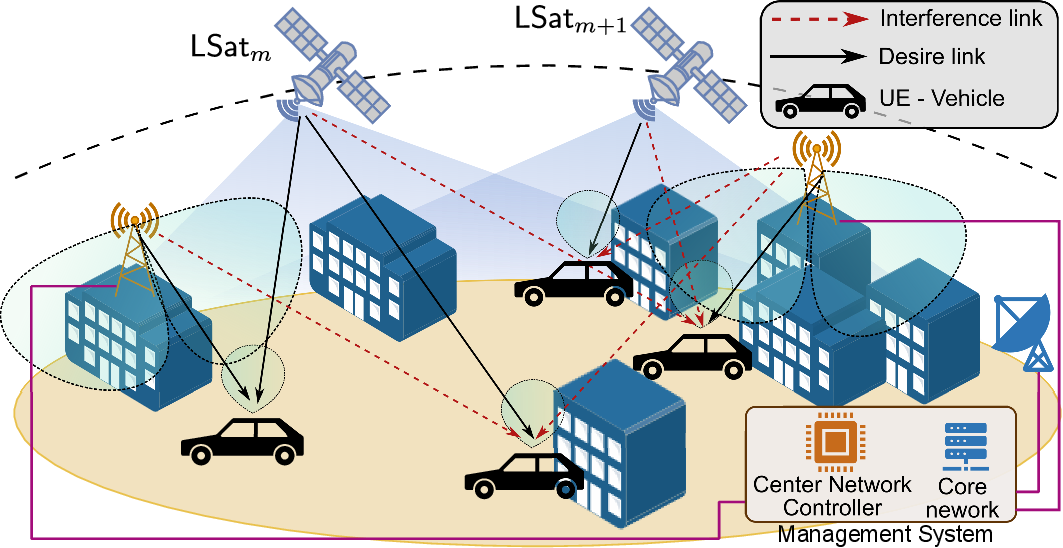}
    \vspace{-1mm}
    \captionsetup{font=small}
    \caption{System model.}
    \label{fig:SystemModel}
    \vspace{-2mm}
\end{figure}

In this work, we study ISTNs as per Fig.~\ref{fig:SystemModel}, where LSats and BSs jointly serve multiple automotive UEs. Specifically, we focus on urban environments where the impact of buildings is considered, and the UEs represent moving vehicles on city streets. According to the current spectrum regulation discussed above with the coexistence and overlap of TN and NTN DL transmission in the C-band, we study the DL operation of ISTNs wherein we assume that TNs and NTNs operate in the same RFB.
% In this work, we study the downlink (DL) operation of ISTNs as per Fig.~\ref{fig:SystemModel}, where LSats and BSs jointly serve multiple automotive UEs. Specifically, we focus on urban environments where the impact of buildings is considered, and the UEs represent moving vehicles on city streets. 
Let $\mathcal{N}\triangleq \{1, \dots,N \},\; \mathcal{M}\triangleq \{1, \dots,M \}$ and $\mathcal{K}\triangleq \{1, \dots,K \}$ denote the sets of BSs, LSats, and UEs, respectively. In addition, let $\sf{BS}_{n},\; \sf{LSat}_{m}$ and $\sf{UE}_{k}$ be the $n-$th BS, $m-$th LSat, and $k-$th UE, respectively. 
To account for other UEs in the system, we also assume the presence of terrestrial UEs (TUEs) and satellite UEs (SUEs), served by BSs and LSats, respectively. Additionally, we employ a window-based model consisting of $N_{\sf{TS}}$ time slots (TSs), where the duration of each TS is $T_{\sf{S}}$. 
During this TW, each automotive vehicle UE is moving in its predetermined route. Subsequently, based on the satellite orbit and UE movement, we assume that the time alignment at LSats is perfect. 
Assume that each LSat operates in a quasi-fixed-beam mode and LSats point their beam towards the designated serving area \cite{3gpp.38.821}.
Additionally, one assumes that BSs and satellite gateways are connected to a management system, wherein each node can exchange data with the core network and the center network controller (CNC) can manage the system resource.

\vspace{-4mm}
\subsection{Tx-UE Channel Model}
\vspace{-1mm}
This subsection presents the channel model of BS-UE and LSat-UE links which follows the coexistence of terrestrial and satellite system scenarios described in 3GPP TR 38863 \cite{3gpp.38.863}.
In particular, for simplification, let $h_{n,k,t}$ and $g_{m,k,t}$ be the effective channel gain coefficients of the ${\sf{BS}}_{n} - {\sf{UE}}_{k}$ and ${\sf{LSat}}_{m} - {\sf{UE}}_{k}$ links at TS $t$ which consist of the path-loss and the radiation beam pattern created by the terminal antennas, respectively. Subsequently, the channel gain is calculated by the inverse of the corresponding path loss (PL). 
% This subsection presents the channel model between one UE and a transmitter side, named Tx which represents both BSs and LSats. Let $h_{n,k,t}$ and $g_{m,k,t}$ be the channel gain of ${\sf{BS}}_{n} - {\sf{UE}}_{k}$ and ${\sf{LSat}}_{m} - {\sf{UE}}_{k}$ links at TS $t$, respectively. Hereafter, the channel gain is calculated by the invert of the corresponding path-loss (PL). 

\subsubsection{Path-loss Model}
Based on the multi-path channel model \cite{Hung_MeditCom24, TWC_Hybrid_Precoding_mMIMO_LEO}, let Tx represent either BS or LSat, the equivalent PL of ${\sf{Tx}}_{x}-{\sf{UE}}_{k}, \; x \in \{m,n\}$ link at TS $t$ can be modeled as
\vspace{-1mm}
\beqn \label{eq: PL model}
\hspace{-7mm} && L_{x,k,t}^{\sf{Tx-UE}} \\
\hspace{-7mm} && = -\log_{10} \scaleobj{0.8}{\left( \left\vert \scaleobj{0.8}{\sum_{i=0}^{N_{\sf{ray}}} }  \frac{G_{\sf{r}}(\varphi_{x,k,i}^{{\sf{a}},t},\theta_{x,k,i}^{{\sf{a}},t}) G_{\sf{t}}(\varphi_{x,k,i}^{{\sf{d}},t},\theta_{x,k,i}^{{\sf{d}},t})}{L_{x,k,i}^{{\sf{pro}},t} L_{x,k}^{{\sf{B}},t} } e^{-j \phi_{x,k,i}^{t}+ j2 \pi t T_{\sf{S}} f_{x,k,i}^{{\sf{D}},t}} \right\vert^{-1} \! \right) }, \nonumber
\eeqn
wherein $N_{\sf{ray}}$ is the number of propagation rays from the Tx to an UE. $L_{x,k,i}^{{\sf{pro}},t}$,  $\phi_{x,k,i}^{t}$, and $f_{x,k,i}^{{\sf{d}},t}$ are the propagation loss, phase delay, and Doppler shift of ray $i$ from ${\sf{Tx}}_{x}$ to ${\sf{UE}}_{k}$ at TS $t$. 
For BS-UE links, we assume the Doppler shift is negligible. For LSat-UE links, due to the very high altitude compared to the distance between UEs and scatterers on the ground, one assumes that the Doppler shifts of all rays are identical, i.e., $f_{m,k,i}^{{\sf{D}},t} = f_{m,k}^{{\sf{D}},t}, \forall m,k,t$ \cite{TWC_Hybrid_Precoding_mMIMO_LEO}. Additionally, thanks to the satellite orbit and UE position knowledge, one assumes that the Doppler shift can be compensated at the LEO satellite payload \cite{3gpp.38.821, Hung_2tierLEO, Yeh_TVT24_DopplerCompensation}. 
Besides, $G_{\sf{r}}(\cdot)$ and $G_{\sf{t}}(\cdot)$ are the antenna gain patterns of the UE and the Tx.
$(\varphi_{x,k,i}^{{\sf{a}},t}, \theta_{x,k,i}^{{\sf{a}},t})$ and $(\varphi_{x,k,i}^{{\sf{d}},t}, \theta_{x,k,i}^{{\sf{d}},t})$ are the elevation and azimuth angles of arrive at ${\sf{UE}}_{k}$ and departure at ${\sf{Tx}}_{x}$ corresponding to ray $i$ at TS $t$. 
Especially, ray $i, \; i \geq 1$, is the LoS ray or the ray reaching UE by reflection or diffraction on obstacles in the environment. In case existing obstacles, ray $i=0$ is the ray propagating through the wall to reach the UE, the channel gain loss of this ray can be modeled as twice of the wall-loss as \cite{3gpp.38.901}
\vspace{-2mm}
\beq
L_{x,k,0}^{{\sf{pro}},t} \! = \! L_{x,k}^{{\sf{FS}},t} \!\! + 2\Big(5 - 10 \log\big(0.7 \cdot 10^{\frac{-L_{\sf{IRRglass}}}{10}} \!\! + 0.3 \cdot 10^{\frac{-L_{\sf{concrete}}}{10}}\big)\Big),
\eeq
$L_{x,k}^{{\sf{FS}},t}$ is the free-space PL, $L_{\sf{IRRglass}}$ and $L_{\sf{concrete}}$ are the penetration loss metrics of IRR glass and concrete materials, respectively. In addition, $L_{x,k}^{{\sf{B}},t}$ is the additional basic loss. For ${\sf{BS}}_{x}-{\sf{UE}}_{k}$ link, since path-loss and shadow fading is accounted in $L_{x,k,i}^{{\sf{pro}},i}$ \cite{3gpp.38.901}, $L_{x,k}^{{\sf{B}},t}=1$. For ${\sf{LSat}}_{x}-{\sf{UE}}_{k}$ link, $L_{x,k}^{{\sf{B}},t}$ is computed by atmospheric absorption and attenuation of rain and clouds \cite{3gpp.38.811}.

\subsubsection{Antenna Model}
\begin{figure}
    \centering
    \begin{subfigure}{0.24\textwidth}
        \centering
        \includegraphics[width=43mm]{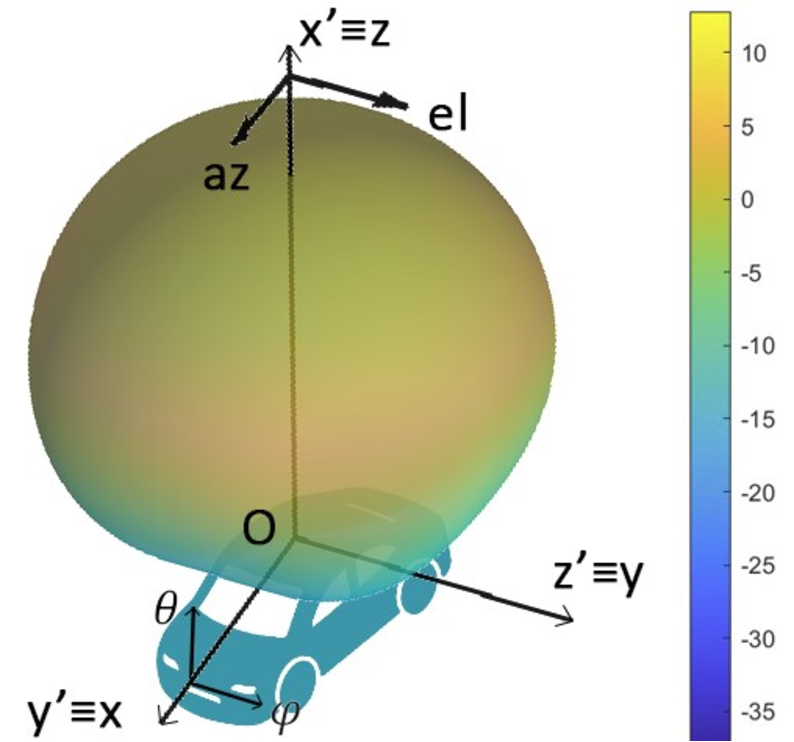}
    \end{subfigure}
    % \hspace{2mm}
    \begin{subfigure}{0.24\textwidth}
        \centering
        \includegraphics[width=43mm]{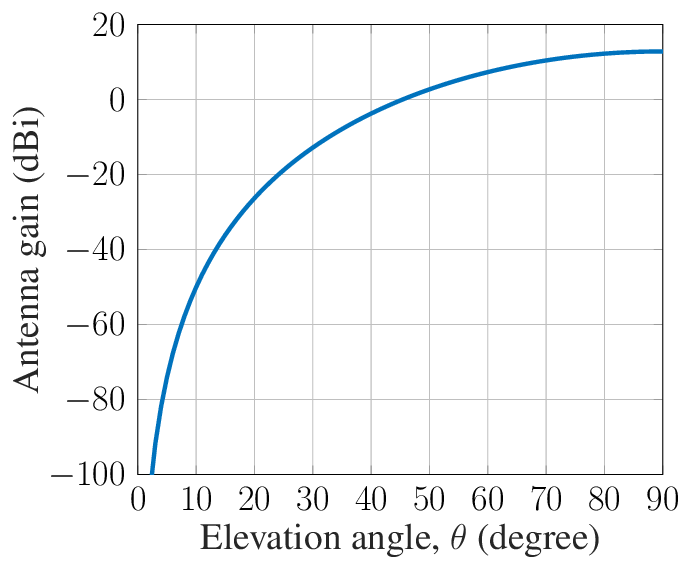}
    \end{subfigure}
    \vspace{-4mm}
    \captionsetup{font=small}
    \caption{The gain pattern of UE's antenna.}
    \label{fig:2x2PatchPattern}
    \vspace{-2mm}
\end{figure}
In this work, we assume that the vehicle UE is equipped with a low-cost patch antenna on the rooftop with no beamforming capabilities, i.e. pointing to the Zenith (as shown in Fig.~\ref{fig:2x2PatchPattern}). Additionally, its radiation pattern can be modeled by employing a cosine pattern as
\vspace{-2mm}
\beq \label{eq: cosine pattern}
    \bar{G}_{\sf{veh}}(az, el) = \cos^{2\bar{m}} (az) \cos^{2 \bar{n}} (el),
\eeq
where $\bar{m}$ and $\bar{n}$ are the order of the. $\sf{az}$ and $\sf{el}$ are the azimuth and elevation angles in the $\sf{Ox'y'z'}$ coordinate system in which its $\sf{Oy'z'}$ plane is mounted on the vehicle's rooftop. Besides, azimuth and elevation angles $\varphi$ and $\theta$ in $G_{\sf{r}}(\varphi, \theta)$ in \eqref{eq: PL model} are measured in the $\sf{Oxyz}$ coordinate system where its $\sf{Oxy}$ plane mounted on the vehicle's rooftop. Hence, it is noted that angles $\sf{az}$ and $\sf{el}$ can be computed from angles $\varphi$ and $\theta$ by the coordinate rotation operation.
For instance, the directivity gain pattern of the patch antenna in the $\sf{Oxyz}$ and $\sf{Ox'y'z'}$ coordinate systems is illustrated in Fig.~\ref{fig:2x2PatchPattern}.
Regarding the LSat's antenna, its antenna gain pattern is modeled as \cite{3gpp.38.821}.
\vspace{-2mm}
\beq
G_{\sf{Sat}}(\theta) = \scaleobj{0.9}{\begin{cases}
	G_{\sf{Sat}}^{\sf{max}}, & \theta = 0, \\
	G_{\sf{Sat}}^{\sf{max}} 4 \left\vert \frac{J_1(k a \sin{\theta})}{ka \sin{\theta}} \right\vert^2, & \theta \neq 0,
\end{cases}}
\eeq
wherein $G_{\sf{S}}^{\sf{max}}$ is the maximum LSat antenna gain, $J_1(\cdot)$ is the Bessel function of the first kind and first order, $k=2\pi f_c / c$. In addition, $a, f_c$ and $ c$ are the antenna aperture radius, operation frequency, and light speed, respectively.
Regarding the BS, for simplicity, we employ the non-active antenna system pattern for the BS which is used in the coexistence scenarios examined in 3GPP TR 38.863 \cite{3gpp.38.863}. Particularly, the beam pattern created by BS's antennas is modeled as
\vspace{-2mm}
\beq
G_{\sf{BS}}(\theta, \varphi) = - \min\{ -(A_{\sf{E,V}}(\theta) + A_{\sf{E,H}}(\varphi)), A_{\sf{m}} \} + G_{\sf{BS}}^{\sf{max}},
\eeq
where $A_{\sf{E,V}}(\theta)$ and $A_{\sf{E,H}}(\varphi)$ are the vertical and horizontal gain patterns which are expressed in \cite{3gpp.38.863}, $G_{\sf{BS}}^{\sf{max}}$ is the maximum gain the the BS's antenna. Besides, a downtill angle of $10^\circ$ is applied to the BS antenna in urban areas \cite{3gpp.38.863}.

\vspace{-3mm}
\subsection{BS/LSat-UE Association Model}
To indicate the BS-UA, we first introduce binary variable $\boldsymbol{\alpha} \triangleq [\alpha_{n,k,t}]_{\forall (n,k,t)}$ as
\vspace{-2mm}
\begin{align}
	\alpha_{n,k,t} = 
	\begin{cases}
		1, & \text{if } {\sf{UE}}_{k} \text{ connects to } {\sf{BS}}_{n} \text{ at TS } t, \\
		0, & \text{otherwise.}
	\end{cases}
\end{align}
In addition, one assumes that each UE can connect to at most one BS at each TS, which is ensured as
\vspace{-1mm}
\beq
(C1): \quad \scaleobj{0.8}{\sum\nolimits_{\forall n}} \alpha_{n, k,t} \leq 1, \forall (k,t).
\eeq
Regarding the available BS-UA at each TS, assuming that ${\sf{BS}}_{n}$ can serve at most $\psi_{n}^{\sf{B}}$ UEs at each TS. Then, the connection limitation at each BS can be cast by 
\vspace{-1mm}
\beq
(C2): \quad \scaleobj{0.8}{\sum\nolimits_{\forall k}} \alpha_{n, k,t} \leq \psi_{n}^{\sf{B}} - \eta_{n,t}^{{\sf{B}}}, \forall (n,t),
\vspace{-1mm}
\eeq
wherein $\eta_{n,t}^{{\sf{B}}}$ is the number of TUEs\footnote{The presence of TUEs is used to emulate the congestion caused by other UEs at BSs.} served by ${\sf{BS}}_{n}$ at TS $t$.
Regarding the LSat-UA, one can see that UEs can connect to the LSat if it is in the field of view (FoV) of UEs. Let $\Theta_{t} = \{\forall (m,k) | \theta_{m,k,t} \geq \tilde{\theta}\}$ be the set of LSat-UE pairs satisfying the FoV criteria above at TS $t$, wherein $\theta_{m,k,t}$ is the elevation angle at ${\sf{UE}}_{k}$ toward ${\sf{LSat}}_{m}$ at TS $t$ and $\tilde{\theta}$ is the minimum elevation angle requirement.
Next, we introduce the LSat-UA variable $\boldsymbol{\beta} \triangleq [\beta_{m,k,t}]_{\forall (m,k,t)}$ corresponding to TS $t$ as 
\vspace{-1mm}
\begin{align} \label{eq: LSat-UA}
	\beta_{m,k,t}\! \! = \! 
	\begin{cases}
		1, & \!\!\! \text{if } (m,k) \! \! \in \! \Theta_{t} \text{ and } {\sf{UE}}_{k} \text{ connects to } {\sf{LSat}}_{m}, \\
		0, & \text{otherwise.}
	\end{cases}
\end{align}
Assuming that each UE can connect to at most one LSat at each TS while each ${\sf{LSat}}_{m}$ can serve at most $\psi_{m}^{\sf{S}}$ UEs at each TS. These yield the following constraints
\vspace{-1mm}
\bieq{ll}
(C3):& \quad \scaleobj{0.8}{\sum\nolimits_{\forall m}} \beta_{m,k,t} \leq 1, \forall (k,t), \\
(C4):& \quad \scaleobj{0.8}{\sum\nolimits_{\forall k}} \beta_{m,k,t}  \leq \psi_{m}^{\sf{S}} - \eta_{m,t}^{{\sf{S}}}, \forall (m,t),
\vspace{-1mm}
\eieq
where $\eta_{m,t}^{{\sf{S}}}$ is the number of SUEs\footnote{The presence of SUEs is used to emulate the congestion caused by other UEs at LSats.} served by ${\sf{LSat}}_{m}$ at TS $t$.

Regarding UE experience, to ensure the quality of experience (QoE) in terms of seamless connectivity and to guarantee that all UEs will be served, we assume that each UE has at least one connection at each TS, which is ensured as
%Furthermore, we assume that dual connectivity is supported \cite{majamaa_ComMag24}, i.e., UE can connect to both BS and LSat simultaneously. To ensure the quality of experience (QoE) in terms of seamless connectivity, one assumes that each UE has at least one connectivity at each TS, which yields the following constraint.
\vspace{-2mm}
\beq \label{eq: QoE const}
(C5): \quad \scaleobj{0.8}{\sum\nolimits_{\forall n}} \alpha_{n,k,t} + \scaleobj{0.8}{\sum\nolimits_{\forall m}} \beta_{m,k,t} \geq 1, \forall (k,t).
\eeq

% Regarding the HO, due to the dynamics in systems, UEs need to change their connections between BS-BS, LSat-LSat, and BS-LSat to improve the rate throughput. 
% Subsequently, based on the UA decision, the CC procedure can be executed depending on the system standard. Thanks to the UA, the number of HOs at each TS can be identified by the difference in the UA between the current and previous TSs. Particularly, the number of HOs in the considered TW can be formed as a function of $\boldsymbol{\alpha}$ and $\boldsymbol{\beta}$ as

Additionally, due to system dynamics, UEs must adapt their association with BSs and LSats to enhance throughput performance. 
Particularly, UEs may carry out the HO, establish an additional connection, or remove the existing connection.
Based on the UA decision, the connection state transition procedure can be executed accordingly to the system standard \cite{3gpp.38.300}. 
% However, these connection state transitions lead to signaling overhead between UEs and the corresponding BS/LSat.
% Furthermore, in the context of dynamic systems, unnecessary connection state changes may arise, which are undesirable and should be minimized. 
Thanks to the UA, the number of connection state changes (CCs) is quantified by evaluating the difference in UA decisions between consecutive TSs as follows
\vspace{-1mm}
\beq \label{eq: fCC}
f^{\sf{CC}}(\boldsymbol{\alpha}, \boldsymbol{\beta}) \\
= \scaleobj{0.8}{\sum_{\forall t>1}} \Big( \scaleobj{0.8}{\sum_{\forall (n,k)}} | \alpha_{n,k,t} - \alpha_{n,k,t-1} | + \!\!\! \scaleobj{0.8}{\sum_{\forall (m,k)}} | \beta_{m,k,t} - \beta_{m,k,t-1}| \Big). 
\vspace{-4mm}
\eeq

\vspace{-6mm}
\subsection{Transmission Model}
\vspace{-1mm}
This subsection describes the transmission from BSs and LSats to UEs. Since TNs and NTNs operate in the same RFB, the transmission involves intra-system interference (IaSI) and inter-system interference (ISI). Depending on the standard in each system, based on the employed resource allocation and multiple access technique, we assume that BSs are coordinated to mitigate the TN IaSI such as in \cite{VuHa_TVT16, AbantoLeon_TWC22, Tam_TWC17} and LSats are coordinated for that in NTNs such as in \cite{VuHa_PIMRC24, XvHongtao_TVT24}. However, due to the long distance between LSats and ground segments, TNs and NTNs are not coordinated to mitigate ISI. Hence, for simplification, we assume that IaSI is negligible and only ISI between TNs and NTNs is considered. 
Consequently, if ${\sf{UE}}_{k}$ connects to ${\sf{BS}}_{n}$, the corresponding signal-to-interference-plus-noise-ratio (SINR) can be expressed as
\vspace{-2mm}
\bieq{ll} \label{eq: SINR_n,k}
\gamma_{n,k,t}^{{\sf{B}}}(\boldsymbol{\alpha}, \boldsymbol{\beta}, \boldsymbol{P}^{\sf{B}}, \boldsymbol{P}^{\sf{S}}) = \scaleobj{0.9}{\frac{\alpha_{n,k,t} P_{n,k,t}^{\sf{B}} h_{n,k,t}}{ \Xi_{k,t}(\boldsymbol{\beta}, \boldsymbol{P}^{\sf{S}}) + \sigma_{k}^2}},
\vspace{-1mm}
\eieq
where $\sigma_{k}^2$ is the additive white Gaussian noise power and $\Xi_{k,t}(\boldsymbol{\beta}, \boldsymbol{P}^{\sf{S}}) = \sum_{\forall m} ( \sum_{\forall k'} \beta_{m,k',t} P_{m,k',t}^{\sf{S}} + \eta_{m,t}^{\sf{S}} \bar{P}_{m,t}^{\sf{S}}) g_{m,k,t} $ is ISI caused by LSats at ${\sf{UE}}_{k}$. 
Specifically, the first summation and the second term in the brackets are the total transmit power of ${\sf{LSat}}_{m}$ to $K$ UEs and its own $\eta_{m,t}^{\sf{S}}$ SUEs, respectively. 
Particularly, $\boldsymbol{P}^{\sf{B}} \triangleq [P_{n,k,t}^{\sf{B}}]_{\forall (n,k,t)}$, $\boldsymbol{P}^{\sf{S}} \triangleq [P_{m,k,t}^{\sf{S}}]_{\forall (m,k,t)}$, 
$P_{n,k,t}^{\sf{B}}$ and $P_{m,k,t}^{{\sf{S}}}$ are the transmit powers of ${\sf{BS}}_{n}$ and ${\sf{LSat}}_{m}$ to ${\sf{UE}}_{k}$ at TS $t$, respectively.  $\bar{P}_{m,t}^{\sf{S}}$ is the transmit power of ${\sf{LSat}}_{m}$ to each of its $\eta_{m,t}^{\sf{S}}$ SUEs which is uniformly allocated as
\vspace{-2mm}
\bieq{ll}
\bar{P}_{m,t}^{\sf{S}} = P_{m}^{\sf{S},max} / \min \{ \psi_{m}^{\sf{S}} , \eta_{m,t}^{\sf{S}} + K \}  , \forall (m,t),
\eieq
wherein $P_{m}^{\sf{S},max}$ is the maximum power budget of ${\sf{LSat}}_{m}$ and $\min \{ \psi_{m}^{\sf{S}}, \eta_{m,t}^{\sf{S}} + K \}$ indicates the maximum number of UEs which can connect to ${\sf{LSat}}_{m}$ at TS $t$. 
Subsequently, the throughput from ${\sf{BS}}_{n}$ to ${\sf{UE}}_{k}$ at TS $t$ can be computed as
\vspace{-3mm}
\beq \label{eq: Rate_n,k}
R_{n,k,t}^{\sf{B}}(\boldsymbol{\alpha}, \boldsymbol{\beta}, \boldsymbol{P}^{\sf{B}}, \boldsymbol{P}^{\sf{S}}) = \log \left(1 + \gamma_{n,k,t}^{\sf{B}}(\boldsymbol{\alpha}, \boldsymbol{\beta}, \boldsymbol{P}^{\sf{B}}, \boldsymbol{P}^{\sf{S}}) \right).
\eeq

Regarding the LSat-UE connection, if ${\sf{UE}}_{k}$ connects to ${\sf{LSat}}_{m}$, 
the corresponding SINR can be expressed as
\vspace{-3mm}
\bieq{ll} \label{eq: SINR_m,k}
\gamma_{m,k,t}^{\sf{S}}(\boldsymbol{\alpha}, \boldsymbol{\beta},  \boldsymbol{P}^{\sf{B}}, \boldsymbol{P}^{\sf{S}}) = \scaleobj{0.9}{\frac{\beta_{m,k,t} P_{m,k,t}^{\sf{S}} g_{m,k,t}}{ \Phi_{k,t}(\boldsymbol{\alpha}, \boldsymbol{P}^{\sf{B}}) + \sigma_{k}^2}}, 
\vspace{-1mm}
\eieq
where $\Phi_{k,t}(\boldsymbol{\alpha}, \boldsymbol{P}^{\sf{B}}) = \sum_{\forall n} \left( \sum_{\forall k'} \alpha_{n,k',t} P_{n,k',t}^{\sf{B}} + \eta_{n,t}^{\sf{B}} \bar{P}_{n,t}^{{\sf{B}}} \right) h_{n,k,t}$ is the ISI caused by BSs, the first summation and the second term inside brackets are the total transmit power of ${\sf{BS}}_{n}$ to $K$ UEs and its own $\eta_{n,t}^{\sf{B}}$ TUEs, respectively. 
Particularly, $\bar{P}_{n,t}^{\sf{B}}$ is the transmit power from ${\sf{BS}}_{n}$ to each of its $\eta_{n,t}^{\sf{B}}$ TUEs which is uniformly allocated as
\vspace{-1mm}
\bieq{ll}
\bar{P}_{n,t}^{\sf{B}} = P_{n}^{\sf{B},max} / \min \{ \psi_{n}^{\sf{B}} , \eta_{n,t}^{\sf{B}} + K \}  , \forall (n,t),
\vspace{-2mm}
\eieq
{$P_{n}^{\sf{B},max}$ is the maximum power budget of ${\sf{BS}}_{n}$ and $\min \{ \psi_{n}^{\sf{B}}, \eta_{n,t}^{\sf{B}} + K \}$ is the maximum number of UEs which can connect to ${\sf{BS}}_{n}$ at TS $t$}. 
Subsequently, the transmission throughput from ${\sf{LSat}}_{m}$ to ${\sf{UE}}_{k}$ can be formulated as
\vspace{-2mm}
\beq \label{eq: Rate_m,k}
R_{m,k,t}^{\sf{S}}(\boldsymbol{\alpha}, \boldsymbol{\beta},  \boldsymbol{P}^{\sf{B}}, \boldsymbol{P}^{\sf{S}}) = \log \left(1 + \gamma_{m,k,t}^{\sf{S}}(\boldsymbol{\alpha}, \boldsymbol{\beta},  \boldsymbol{P}^{\sf{B}}, \boldsymbol{P}^{\sf{S}}) \right).
\vspace{-1mm}
\eeq
Hence, the aggregated transmission throughput of ${\sf{UE}}_{k}$ at TS $t$ can be expressed as
\vspace{-1mm}
\beq
R_{k,t} \!(\boldsymbol{\alpha}, \boldsymbol{\beta},  \boldsymbol{P}^{\sf{B}} \!\!, \! \boldsymbol{P}^{\sf{S}}) \!=\! 
\scaleobj{0.8}{\sum_{\forall n}} R_{n,k,t}^{\sf{B}}\!(\boldsymbol{\alpha}, \boldsymbol{\beta},  \boldsymbol{P}^{\sf{B}} \!\!,\! \boldsymbol{P}^{\sf{S}}) 
%\nonumber \\ \hspace{35mm} 
\!+\! \scaleobj{0.8}{\sum_{\forall m}} R_{m,k,t}^{\sf{S}}\!(\boldsymbol{\alpha}, \boldsymbol{\beta},  \boldsymbol{P}^{\sf{B}} \!\!,\! \boldsymbol{P}^{\sf{S}}). \vspace{-1mm}
\eeq

Regarding the QoS of UEs, the rate threshold constraint should be considered. However, ensuring this constraint at each TS may be infeasible since the transmission links can be blocked by buildings in urban environments at certain TSs. Hence, one can guarantee the QoS of UEs by considering the minimum average rate constraint at consecutive periods as
\vspace{-1mm}
\bieq{ll}
    (C6): \quad  \scaleobj{0.75}{\frac{1}{N_{\sf{TS}}^{\sf{QoS}}} } \scaleobj{0.8}{\sum_{\forall t \in \mathcal{T}_{q} } } R_{k,t}(\boldsymbol{\alpha}, \boldsymbol{\beta},  \boldsymbol{P}^{\sf{B}}, \boldsymbol{P}^{\sf{S}}) \geq \bar{R}_{k}, \forall k, q > 0, \nonumber
    \vspace{-1mm}
\eieq
wherein $\mathcal{T}_{q} = \{ (q-1) N_{\sf{TS}}^{\sf{QoS}}+1,\dots, q N_{\sf{TS}}^{\sf{QoS}} \}$ is the set of TSs in  period $q$, each period consists of $N_{\sf{TS}}^{\sf{QoS}}$ TSs, and $\bar{R}_{k}$ is the average rate threshold of ${\sf{UE}}_{k}$ at each period.

\vspace{-3mm}
\subsection{Problem Formulation}
\vspace{-1mm}
In this subsection, we formulate a multi-objective optimization problem aiming to maximize the average system SR and minimize the average number of CCs per TS, wherein the UA and PA are jointly optimized. 

Assuming that the deployment of BSs and LSats is sufficiently dense, UEs may have multiple connection options at each TS. However, due to the limitation of radio resources, BSs and LSats can overload at certain times depending on the number of connected devices. In addition, the system is highly dynamic due to UE mobility, LSat movement, and the complexity of urban environments, leading to frequent variations in link availability and quality over time.
 
Therefore, to ensure connectivity under such dynamics, UEs must appropriately adapt their associations with BSs and LSats through HO, establishing new connections, or releasing existing ones.
Concurrently, due to the BS/LSat's transmit power limit and the ISI impact, an appropriate PA is essential.

Besides, each connection state transition incurs signaling overhead between UEs and the corresponding BS/LSat. Furthermore, unnecessary state transitions may arise, which are undesirable and should be minimized. 
Therefore, in addition to enhancing throughput and enabling seamless HO, the optimization goal must also aim to minimize the number of CCs. Particularly, the optimization is mathematically formulated as 
\bieq {rl}\label{eq: SRCC Prob 1}
(\mathcal{P}_{0}) 
\max_{\boldsymbol{\alpha}, \boldsymbol{\beta},  \boldsymbol{P}^{\sf{B}}, \boldsymbol{P}^{\sf{S}}} & \; \scaleobj{0.8}{\frac{\rho}{N_{\sf{TS}}} } \scaleobj{0.8}{\sum_{\forall (k, t)}} R_{k,t}(\boldsymbol{\alpha}, \boldsymbol{\beta},  \boldsymbol{P}^{\sf{B}}, \boldsymbol{P}^{\sf{S}}) - \scaleobj{0.8}{\frac{1-\rho}{N_{\sf{TS}}}} f^{\sf{CC}}(\boldsymbol{\alpha}, \boldsymbol{\beta}) \nonumber \\
 \st \quad & \text{constraints } (C1)-(C6), \nonumber \\
 (C0): \quad & \alpha_{n,k,t}, \beta_{m,k,t} \in \{0,1\} , \forall (m,n,k,t), \nonumber \\
 (C7): \quad & \eta_{n,t}^{\sf{B}} \bar{P}_{n,t}^{{\sf{B}}} + \scaleobj{0.8}{\sum\nolimits_{\forall k}} \alpha_{n,k,t} P_{n,k,t}^{\sf{B}} \leq P_{n}^{\sf{B,max}},  \forall (n,t), \nonumber \\
(C8): \quad  & \eta_{m,t}^{\sf{S}} \bar{P}_{m,t}^{{\sf{S}}} + \scaleobj{0.8}{\sum\nolimits_{\forall k}} \beta_{m,k,t} P_{m,k,t}^{\sf{S}} \leq P_{m}^{\sf{S,max}}, \forall (m,t), \nonumber
\eieq
wherein $\rho \in (0,1)$ is the multi-objective factor, a higher $\rho$ leads to higher priority of SR maximization. In addition, constraints $(C7)$ and $(C8)$ ensure that the transmit power does not exceed the BS and LSat power budgets, respectively. The outcome of solving this problem can provide solutions for UA and PA over time, helping to improve system performance. However, due to the non-convexity of the rate functions, and the coupling between binary and continuous variables, this problem belongs to mixed-integer non-linear programming which is difficult to solve directly. 

\vspace{-2mm}
\section{Proposed Solutions} \label{sec: propose sol}
\begin{figure}
    \centering
    \includegraphics[width=85mm]{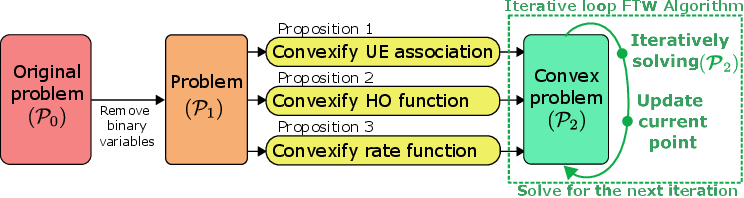}
    \vspace{-1mm}
    \captionsetup{font=small}
    \caption{Flowchart of the full-time-window algorithm}
    \label{fig:flowchart_fullwindowAlg}
    \vspace{-2mm}
\end{figure}

\vspace{-1mm}
This section proposed two algorithms for the full time window and for the prediction time window. The developing procedure of the first algorithm is summarized in Fig.~\ref{fig:flowchart_fullwindowAlg}.

\vspace{-4mm}
\subsection{Problem Transformation}
Analyzing the $(\mathcal{P}_{0})$ structure and its optimization variables, we obtain the following relationships, each TS $t$
\begin{itemize}
	\item If ${\sf{UE}}_{k}$ connects to ${\sf{BS}}_{n}$, $P_{n,k,t}^{\sf{B}} > 0$, and $P_{n,k,t}^{\sf{B}} = 0$ if ${\sf{UE}}_{k}$ does not connect to ${\sf{BS}}_{n}$.
	\item If ${\sf{UE}}_{k}$ connects to ${\sf{LSat}}_{m}$, $P_{m,k,t}^{\sf{S}} > 0$, and $P_{m,k,t}^{\sf{S}} = 0$ if ${\sf{UE}}_{k}$ is not served by ${\sf{LSat}}_{m}$.
\end{itemize}
Based on these relations, the binary variable can be represented by the $\ell_0-$norm components of continuous ones as
\vspace{-1mm}
\bieq{ll}\label{eq: binary reduce}
	\alpha_{n,k,t} = \Vert P_{n,k,t}^{\sf{B}} \Vert_{0}, \forall (n,k,t), \subnum \\
	\beta_{m,k,t} = \Vert P_{m,k,t}^{\sf{S}} \Vert_{0}, \forall (m,k,t). \subnum
\eieq
One notes that representations in \eqref{eq: binary reduce} are valid for $\forall P_{n,k,t}^{\sf{B}} \geq 0$ and $\forall P_{m,k,t}^{\sf{S}} \geq 0$.
Additionally, due to the FoV criteria for LSat-UA in \eqref{eq: LSat-UA}, the LSat's power control elements are set as $P_{m,k,t}^{\sf{S}} = 0, \; \forall (m,k) \notin \Theta_{t}$. 
By employing the relations in \eqref{eq: binary reduce}, the binary terms in rate and SINR functions can be omitted. Subsequently, problem $\mathcal{P}_{0}$ can be rewritten as
\bieq {lrl}\label{eq: SRCC Prob 2}
(\mathcal{P}_{1}): 
& \max_{\boldsymbol{P}^{\sf{B}}, \boldsymbol{P}^{\sf{S}}, \boldsymbol{\lambda}^{\sf{B}},  \boldsymbol{\lambda}^{\sf{S}}} \quad & \! \scaleobj{0.8}{\frac{\rho}{N_{\sf{TS}}}} \tilde{\lambda} - \scaleobj{0.8}{\frac{1-\rho}{N_{\sf{TS}}}} \bar{f}^{\sf{CC}}(\boldsymbol{P}^{\sf{B}}, \boldsymbol{P}^{\sf{S}}) \nonumber \\
&\st \quad
(\bar{C}1): \quad & \! \scaleobj{0.8}{\sum_{\forall n}} \Vert P_{n,k,t}^{\sf{B}} \Vert_{0} \leq 1, \forall (k,t), \nonumber \\
&(\bar{C}2): \quad & \! \scaleobj{0.8}{\sum_{\forall k}} \Vert P_{n,k,t}^{\sf{B}} \Vert_{0} \leq \psi_{n}^{\sf{B}} - \eta_{n,t}^{\sf{B}}, \forall (n,t), \nonumber \\
&(\bar{C}3): \quad & \! \scaleobj{0.8}{\sum_{\forall m}} \Vert P_{m,k,t}^{\sf{S}} \Vert_{0} \leq 1, \forall (k,t), \nonumber \\
&(\bar{C}4): \quad &\! \scaleobj{0.8}{\sum_{\forall k}} \Vert P_{m,k,t}^{\sf{S}} \Vert_{0}  \leq \psi_{m}^{\sf{S}} - \eta_{m,t}^{\sf{S}}, \forall (m,t), \nonumber \\
&(\bar{C}5): \quad & \! \scaleobj{0.8}{\sum_{\forall n}} \Vert P_{n,k,t}^{\sf{B}} \Vert_{0} + \scaleobj{0.8}{\sum_{\forall m}} \Vert P_{m,k,t}^{\sf{S}} \Vert_{0} \geq 1, \forall (k,t), \nonumber \\
& (\tilde{C}6): \quad & \! \scaleobj{0.8}{\frac{1}{N_{\sf{TS}}^{\sf{QoS}}}} \scaleobj{0.8}{\sum_{\forall t \in \mathcal{T}_{q}}} (\scaleobj{0.8}{\sum_{\forall n}} \lambda_{n,k,t}^{\sf{B}} + \scaleobj{0.8}{\sum_{\forall m}} \lambda_{m,k,t}^{\sf{S}}) \geq \bar{R}_{k} , \forall (k,q), \nonumber \\
&(\tilde{C}7): \quad &\eta_{n,t}^{\sf{B}} \bar{P}_{n,t}^{\sf{B}} + \scaleobj{0.8}{\sum_{\forall k}} P_{n,k,t}^{\sf{B}} \leq P_{n}^{\sf{B,max}}, \forall (n,t), \nonumber \\
&(\tilde{C}8): \quad &\eta_{m,t}^{\sf{S}} \bar{P}_{m,t}^{\sf{S}} + \scaleobj{0.8}{\sum_{\forall k}} P_{m,k,t}^{\sf{S}} \leq P_{m}^{\sf{S,max}}, \forall (m,t), \nonumber \\
&(C9):\quad &\lambda_{n,k,t}^{\sf{B}} \leq R_{n,k,t}^{\sf{B}}(\boldsymbol{P}^{\sf{B}}, \boldsymbol{P}^{\sf{S}}), \quad \forall (n,k,t), \nonumber \\
&(C10):\quad &\lambda_{m,k,t}^{\sf{S}} \leq R_{m,k,t}^{\sf{S}}(\boldsymbol{P}^{\sf{B}}, \boldsymbol{P}^{\sf{S}}), \quad \forall (m,k,t), \nonumber
\eieq
where $\bar{f}^{\sf{CC}}(\boldsymbol{P}^{\sf{B}}, \boldsymbol{P}^{\sf{S}})$ is rewritten from $f^{\sf{CC}}(\boldsymbol{\alpha}, \boldsymbol{\beta})$ as
\bieq{ll} \label{eq: fCC 2}
\bar{f}^{\sf{CC}}(\boldsymbol{P}^{\sf{B}}, \boldsymbol{P}^{\sf{S}}) = \scaleobj{0.8}{\sum_{\forall t > 1}}( \scaleobj{0.8}{\sum_{\forall (n,k)}} \left| \Vert P_{n,k,t}^{\sf{B}} \Vert_{0} - \Vert P_{n,k,t-1}^{\sf{B}} \Vert_{0} \right| \nonumber \\
\hspace{25mm} + \scaleobj{0.8}{\sum_{\forall (m,k)}} \left| \Vert P_{m,k,t}^{\sf{S}} \Vert_{0} - \Vert P_{m,k,t-1}^{\sf{S}} \Vert_{0} \right| ). 
\eieq
In addition, $\tilde{\lambda} = \sum_{n,k,t} \lambda_{n,k,t}^{\sf{B}} + \sum_{\forall m,k,t} \lambda_{m,k,t}^{\sf{S}} $, and $\boldsymbol{\lambda}^{\sf{B}} \triangleq [\lambda_{n,k,t}^{\sf{B}}]_{\forall (n,k,t)}$ and $  \boldsymbol{\lambda}^{\sf{S}} \triangleq [\lambda_{m,k,t}^{\sf{S}}]_{\forall (m,k,t)} $ are the slack variables which are considered the lower bounds of the rate functions as in constraints $(C9)$ and $(C10)$, respectively. 
It can be seen that problem $(\mathcal{P}_{1})$  is still difficult to solve due to the nonconvex rate constraint $(C9)-(C10)$ and the sparse $\ell_{0}-$norm terms. 
The sparsity in $\ell_{0}-$norm terms and the non-convexity in rate constraints will be addressed by appropriate approaches in the following sub-sections.

\subsection{Proposed Iterative Solution}
\subsubsection{Approximate $\ell_0$-norm components}
To address the sparsity in problem $(\mathcal{P}_{1})$, the $\ell_{0}$-norm components can be approximated by a concave function as
\bieq{ll} \label{eq: norm0 apx}
	\Vert P_{n,k,t}^{\sf{B}} \Vert_{0} \approx f_{\sf{apx}}(P_{n,k,t}^{\sf{B}}), \forall (n,k,t), \subnum \\
	\Vert P_{m,k,t}^{\sf{S}} \Vert_{0} \approx f_{\sf{apx}}(P_{m,k,t}^{\sf{S}}), \forall (m,k,t), \subnum
\eieq
in which $f_{\sf{apx}}(x) \triangleq 1 - e^{-\zeta x},\; x \geq 0 $ is the approximate concave function, $\zeta$ is a positive number used to control the smoothness of the approximated function \cite{VuHa_TVT16}. Thanks to \eqref{eq: norm0 apx}, $(\bar{C}5)$ can be transformed into the convex constraint as
\beq
    (\tilde{C}5): \quad \scaleobj{0.8}{\sum_{\forall n}} f_{\sf{apx}}(P_{n,k,t}^{\sf{B}}) + \scaleobj{0.8}{\sum_{\forall m}} f_{\sf{apx}}(P_{m,k,t}^{\sf{S}}) \geq 1, \forall (k,t). \nonumber
\eeq

\begin{proposition} \label{pro: up bound norm0 const}
    Constraints $(\bar{C}1)-(\bar{C}4)$ can be approximated respectively at iteration $(i+1)$ as
    \bieq{ll}
        (\tilde{C}1): \quad \scaleobj{0.8}{\sum_{\forall n}} \tilde{f}_{\sf{apx}}^{(i)}(P_{n,k,t}^{\sf{B}}) \leq 1, \forall (k,t), \nonumber \\
        (\tilde{C}2): \quad \scaleobj{0.8}{\sum_{\forall k}} \tilde{f}_{\sf{apx}}^{(i)}(P_{n,k,t}^{\sf{B}}) \leq \psi_{n}^{\sf{B}} - \eta_{n,t}^{\sf{B}}, \forall (n,t), \nonumber \\
        (\tilde{C}3): \quad \scaleobj{0.8}{\sum_{\forall m}} \tilde{f}_{\sf{apx}}^{(i)}(P_{m,k,t}^{\sf{S}}) \leq 1, \forall (k,t), \nonumber \\
        (\tilde{C}4): \quad \scaleobj{0.8}{\sum_{\forall k}} \tilde{f}_{\sf{apx}}^{(i)}(P_{m,k,t}^{\sf{S}})  \leq \psi_{m}^{\sf{S}} - \eta_{m,t}^{\sf{S}}, \forall (m,t), \nonumber
    \eieq
    wherein $\tilde{f}_{\sf{apx}}^{(i)}(x) \triangleq \zeta \exp(-\zeta x^{(i)}) (x - x^{(i)}) - \exp(-\zeta x^{(i)}) + 1 $, $x^{(i)}$ is the feasible point at iteration $i$.
\end{proposition}
\begin{IEEEproof}
    Thanks to \eqref{eq: norm0 apx}, the $\ell_{0}$-norm components in constraints $(\bar{C}1) - (\bar{C}4)$ can be approximated by $f_{\sf{apx}}(\cdot)$. However, the approximated constraints are nonconvex due to the concavity of $f_{\sf{apx}}(\cdot)$. 
    Subsequently, by utilizing the first-order approximation, an upper bound of $f_{\sf{apx}}(x)$ around $x^{(i)}$ is obtained as
    \beq \label{eq: up bound fapx}
        f_{\sf{apx}}(x) \leq \nabla f_{\sf{apx}}(x) |_{x=x^{(i)}} (x - x^{(i)}) + f_{\sf{apx}}(x^{(i)}) := \tilde{f}_{\sf{apx}}^{(i)}(x).
    \eeq
    By employing \eqref{eq: up bound fapx}, constraints $(\bar{C}1) - (\bar{C}4)$ are convexified as $(\tilde{C}1) - (\tilde{C}4)$, respectively.
\end{IEEEproof}

\begin{proposition} \label{pro: up bound fCC}
    An upper bound of the CC-number function can be expressed as
    \bieq{ll} \label{eq: fCC 3}
    \bar{f}^{\sf{CC}}(\boldsymbol{P}^{\sf{B}}, \boldsymbol{P}^{\sf{S}}) \leq \scaleobj{0.8}{\sum_{\forall t>1}} ( \scaleobj{0.8}{\sum_{\forall (n,k)}} (a_{n,k,t}^{\sf{up}} - a_{n,k,t}^{\sf{low}})  \nonumber \\
    \hspace{10mm} + \scaleobj{0.8}{\sum_{\forall (m,k)}} (b_{m,k,t}^{\sf{up}} - b_{m,k,t}^{\sf{low}}) ) 
    := \tilde{f}^{{\sf{CC}},(i)}(\boldsymbol{a}, \boldsymbol{b}),
    \eieq
    where $\boldsymbol{a} = [a_{n,k,t}^{\sf{low}},a_{n,k,t}^{\sf{up}}]_{\forall (n,k,t)}$ and $\boldsymbol{b} = [b_{m,k,t}^{\sf{low}},b_{m,k,t}^{\sf{up}}]_{\forall (m,k,t)}$ are the new slack variables satisfying constraints $\forall (m,n,k,t)$:
    \bieq{ll}
        (\tilde{C}11.1): a_{n,k,t}^{\sf{low}} \leq f_{\sf{apx}}(P_{n,k,t}^{\sf{B}}), 
        \; a_{n,k,t}^{\sf{low}} \leq f_{\sf{apx}}(P_{n,k,t-1}^{\sf{B}}), \nonumber \\
        (\tilde{C}11.2): a_{n,k,t}^{\sf{up}} \geq f_{\sf{apx}}^{(i)}(P_{n,k,t}^{\sf{B}}), 
        \; a_{n,k,t}^{\sf{up}} \geq f_{\sf{apx}}^{(i)}(P_{n,k,t-1}^{\sf{B}}), \nonumber \\
        (\tilde{C}12.1): b_{m,k,t}^{\sf{low}} \leq f_{\sf{apx}}(P_{m,k,t}^{\sf{S}}), 
        \; b_{m,k,t}^{\sf{low}} \leq f_{\sf{apx}}(P_{m,k,t-1}^{\sf{S}}),  \nonumber \\
        (\tilde{C}12.2): b_{m,k,t}^{\sf{up}} \geq f_{\sf{apx}}^{(i)}(P_{m,k,t}^{\sf{S}}), 
        \; b_{m,k,t}^{\sf{up}} \geq f_{\sf{apx}}^{(i)}(P_{m,k,t-1}^{\sf{S}}). \nonumber
    \eieq   
\end{proposition}

\begin{IEEEproof}
    The proof is given in Appendix~\ref{app: up bound fCC}.
\end{IEEEproof}

\subsubsection{Convexify rate constraints}
Rate constraints $(C9)$ and $(C10)$ can be convexified by the following proposition.
\begin{proposition} \label{pro: convexify rate}
	Constraints $(C9),(C10)$ can be convexified as
	\bieq{ll} \label{eq: convexify rate}
	(\tilde{C}9.1) \!\!: \ln( P_{n,k,t}^{\sf{B}} h_{n,k,t} \!+\! \Xi_{k,t}(\boldsymbol{P}^{\sf{S}}) \!+\! \sigma_{k}^2)  \! \geq \! \lambda_{n,k,t}^{\sf{B}} \!+\! \mu_{k,t}^{\sf{B}}, \; \forall (n,k,t), \quad \nonumber \\
	(\tilde{C}9.2) \!\! : \Xi_{k,t}(\boldsymbol{P}^{\sf{S}}) \!+\! \sigma_{k}^2 \leq \exp(\mu_{k,t}^{{\sf{B}},(i)})(\mu_{k,t}^{\sf{B}} \!-\! \mu_{k,t}^{{\sf{B}},(i)} \!+\! 1),\; \forall (k,t), \quad  \nonumber \\
	(\tilde{C}10.1)\!\!: \ln(P_{m,k,t}^{\sf{S}} g_{m,k,t} \!+\! \Phi_{k,t}(\boldsymbol{P}^{\sf{B}}) \!+\! \sigma_{k}^2)  \!\!\geq\!\! \lambda_{m,k,t}^{\sf{S}} \!+\! \mu_{k,t}^{\sf{S}},  \forall (m,k,t), \quad \nonumber \\
	(\tilde{C}10.2)\!\!: \Phi_{k,t}(\boldsymbol{P}^{\sf{B}}) \!+\! \sigma_{k}^2 \leq \exp(\mu_{k,t}^{{\sf{S}},(i)})(\mu_{k,t}^{\sf{S}} \!-\! \mu_{k,t}^{{\sf{S}},(i)} \!+\! 1),\; \forall (k,t), \quad  \nonumber
	\eieq
\end{proposition}
\begin{IEEEproof}
    The proof is given in Appendix~\ref{app: convexify rate}
\end{IEEEproof}

Thanks to Propositions~\ref{pro: up bound norm0 const}, \ref{pro: up bound fCC}, and \ref{pro: convexify rate}, problem $(\mathcal{P}_{1})$ can be transformed into a successive convex problem at iteration $i$ as
\bieq {lll}\label{eq: SRCC Prob convex}
(\mathcal{P}_{2}): &
\max_{\substack{ \boldsymbol{P}^{\sf{B}}, \boldsymbol{P}^{\sf{S}}, \boldsymbol{\lambda}^{\sf{B}},  \boldsymbol{\lambda}^{\sf{S}}, \\ \boldsymbol{\mu}^{\sf{B}}, \boldsymbol{\mu}^{\sf{S}} , \boldsymbol{a}, \boldsymbol{b} }}  & \quad  \frac{\rho}{N_{\sf{TS}}} \tilde{\lambda} - \frac{1 - \rho}{N_{\sf{TS}}} \tilde{f}^{{\sf{CC}},(i)}(\boldsymbol{a}, \boldsymbol{b}) \nonumber \\
&& \st \quad (\tilde{C}1)-(\tilde{C}12), \nonumber
\eieq
with $\boldsymbol{\mu}^{\sf{B}} = [\mu_{n,k,t}^{\sf{B}}]_{\forall (n,k,t)}$ and $\boldsymbol{\mu}^{\sf{S}} = [\mu_{m,k,t}^{\sf{S}}]_{\forall (m,k,t)}$.
By iteratively solving problem $(\mathcal{P}_{2})$, we obtain the PA solution for problem $(\mathcal{P}_{0})$. Besides, the UA solution is retrieved as 
%\cite{VuHa_TVT16}
\bieq{ll} \label{eq: recover binary}
    \alpha_{n,k,t} \!=\! 
    \begin{cases}
        1, \text{ if } P_{n,k,t}^{\sf{B}} \! \geq \! \epsilon, \\
        0, \text{ in otherwise },
    \end{cases} \hspace{-2mm} 
    ,\beta_{m,k,t} = 
    \begin{cases}
        1, \text{ if } P_{m,k,t}^{\sf{S}} \! \geq \! \epsilon, \\
        0, \text{ in otherwise },
    \end{cases} 
\eieq
wherein $\epsilon$ is a small positive number. The proposed iterative algorithm for full time-window-based (FTW), so called FTW algorithm, is summarized in Algorithm~\ref{alg: OpzPower FullTS}. 
The convergence of this algorithm is discussed in the following proposition.
\begin{proposition} \label{pro: FullWAlg converg}
    Algorithm~\ref{alg: OpzPower FullTS} converges to a local optimal solution of problem $(\mathcal{P}_{0})$.
    % The sequence of the objective values generated by Algorithm~\ref{alg: OpzPower FullTS} is non-decreasing and converges.
\end{proposition}
\begin{IEEEproof} \label{app: proof FullWAlg converg}
    First, one needs to prove the convergence of Algorithm~\ref{alg: OpzPower FullTS}. Let $\Omega \triangleq (\boldsymbol{P}^{\sf{B}}, \boldsymbol{P}^{\sf{S}}, \boldsymbol{\lambda}^{\sf{B}},  \boldsymbol{\lambda}^{\sf{S}}, \boldsymbol{\mu}^{\sf{B}}, \boldsymbol{\mu}^{\sf{S}}, \boldsymbol{a}, \boldsymbol{b} )$ be the variable tuple, and $\Omega^{(i)}$, $\Omega^{\star (i)}$, and $\mathcal{F}^{(i)}$ be the feasible point, optimal solution, and feasible set of $(\mathcal{P}_{2})$ at iteration $i$.
    In addition, let $F^{(i)}(\Omega)$ denote the objective value of $(\mathcal{P}_{2})$ with solution $\Omega$ at iteration $i$ and feasible point $\Omega^{(i)}$.
    Since $(\mathcal{P}_{2})$ is derived by employing the SCA technique, it satisfies the properties of SCA-based problem \cite{SCA}. In particular, we have $\Omega^{\star (i)} \in \mathcal{F}^{(i)} \cap \mathcal{F}^{(i+1)} $. This leads to $F^{(i+1)}(\Omega^{\star (i+1)}) \geq F^{(i)}(\Omega^{\star (i)})$. Hence, the objective value sequence is non-decreasing. Additionally, the transmit power is bounded by constraints $(\bar{C}7)-(\bar{C}8)$, which prove the convergence of Algorithm~\ref{alg: OpzPower FullTS}. Subsequently, with $(\boldsymbol{\alpha},\boldsymbol{\beta})$ recovered from $(\boldsymbol{P}^{\sf{B}},\boldsymbol{P}^{\sf{S}})$ by \eqref{eq: recover binary}, one can see that the feasible set of $(\mathcal{P}_{2})$ is subset of that of $(\mathcal{P}_{0})$. Therefore, Algorithm~\ref{alg: OpzPower FullTS} converges to a local solution of $(\mathcal{P}_{0})$.
\end{IEEEproof}

% \begin{proposition} \label{pro: FullWAlg converg}
%     The sequence of the objective values generated by Algorithm~\ref{alg: OpzPower FullTS} is non-decreasing and converges.
% \end{proposition}
% \begin{IEEEproof} \label{app: proof FullWAlg converg}
%     First, let $\Omega \triangleq (\boldsymbol{P}^{\sf{B}}, \boldsymbol{P}^{\sf{S}}, \boldsymbol{\lambda}^{\sf{B}},  \boldsymbol{\lambda}^{\sf{S}}, \boldsymbol{\mu}^{\sf{B}}, \boldsymbol{\mu}^{\sf{S}}, \boldsymbol{a}, \boldsymbol{b} )$ be the tuple of all variable, and $\Omega^{(i)}$, $\Omega^{\star (i)}$, and $\mathcal{F}^{(i)}$ be the feasible point, optimal solution, and feasible set of problem $(\mathcal{P}_{2})$ at iteration $i$. In addition, let $F^{(i)}(\Omega)$ denote the objective value with solution $\Omega$ of $(\mathcal{P}_{2})$ at iteration $i$ and feasible point $\Omega^{(i)}$.
%     Since problem $(\mathcal{P}_{2})$ is derived by employing the SCA technique, it satisfies the properties of SCA-based problem \cite{SCA}. In particular, we have $\Omega^{\star (i)} \in \mathcal{F}^{(i)} \cap \mathcal{F}^{(i+1)} $. This leads to $F^{(i+1)}(\Omega^{\star (i+1)}) \geq F^{(i)}(\Omega^{\star (i)})$. Therefore, the sequence of objective values is non-decreasing. In addition, the transmit power is bounded by constraints $(\bar{C}6)-(\bar{C}7)$ leading to convergence of the objective value sequence. This completes the proof of proposition~\ref{pro: FullWAlg converg}.
% \end{IEEEproof}

To solve $(\mathcal{P}_{2})$, one requires an initial point $(\boldsymbol{P}^{{\sf{B}}, {(0)}}, \boldsymbol{P}^{{\sf{S}}, {(0)}}, \boldsymbol{\mu}^{{\sf{B}}, {(0)}}, \boldsymbol{\mu}^{{\sf{S}}, {(0)}})$ satisfying $(\tilde{C}1)-(\tilde{C}12)$. To obtain an initial point, we can relax and penalize constraints $(\tilde{C}5)$, $(\tilde{C}6)$ \cite{Hung_Access22}, and solve the following problem 
\bieq {lll}\label{eq: init point SRCC}
    \min_{\substack{ \boldsymbol{P}^{\sf{B}}, \boldsymbol{P}^{\sf{S}}, \boldsymbol{\lambda}^{\sf{B}},  \boldsymbol{\lambda}^{\sf{S}}, \\ \boldsymbol{\mu}^{\sf{B}}, \boldsymbol{\mu}^{\sf{S}} , \boldsymbol{\tau}, \boldsymbol{\chi} }}  \quad \sum \tau_{k,t} + \sum \chi_{k,q}  \subnum \\
    \st \quad (\tilde{C}1)-(\tilde{C}4),(\tilde{C}7)-(\tilde{C}10), \nonumber \\
    \quad  \! \scaleobj{0.8}{\sum_{\forall n}} f_{\sf{apx}}(P_{n,k,t}^{\sf{B}}) + \scaleobj{0.8}{\sum_{\forall m}} f_{\sf{apx}}(P_{m,k,t}^{\sf{S}}) + \tau_{k,t} \geq 1, \forall (k,t), \subnum \\
    \quad  \! \scaleobj{0.8}{\frac{1}{N_{\sf{TS}}^{\sf{QoS}}}} \scaleobj{0.8}{\sum_{\forall t \in \mathcal{T}_{q}}} (\scaleobj{0.8}{\sum_{\forall n}} \lambda_{n,k,t}^{\sf{B}} + \scaleobj{0.8}{\sum_{\forall m}} \lambda_{m,k,t}^{\sf{S}}) + \chi_{k,q} \geq \bar{R}_{k} , \forall (k,q), \subnum \\
    \quad \tau_{k,t} \geq 0, \chi_{k,q} \geq 0, \; \forall (k,t,q) \subnum
\eieq
where $\boldsymbol{\tau}=\{\tau_{k,t}\}_{\forall (k,t)}$ and $\boldsymbol{\chi}=\{\chi_{k,q}\}_{\forall (k,q)}$ are slack variables. An initial point can be found by solving problem \eqref{eq: init point SRCC} until convergence.

\begin{algorithm}[!t]
\footnotesize
	\begin{algorithmic}[1]
 \captionsetup{font=small}
		\protect\caption{\textsc{FTW Algorithm for Solving Problem $(\mathcal{P}_{0})$}}
		\label{alg: OpzPower FullTS}
		% \long\def\algorithmicrequire{\textbf{Phase 1:}}
		% \REQUIRE
        % \STATE Set $t=1$.\\
        % \WHILE {$t \leq N_T$ or $d_k[t]=0, \forall k$}
        \STATE Set $i=0$ and generate an initial point $(\boldsymbol{P}^{{\sf{B}}, {(0)}}, \boldsymbol{P}^{{\sf{S}}, {(0)}}, \boldsymbol{\mu}^{{\sf{B}}, {(0)}}, \boldsymbol{\mu}^{{\sf{S}}, {(0)}})$ by solving problem \eqref{eq: init point SRCC}.\\
		\REPEAT
		\STATE Solve problem $(\mathcal{P}_{2})$ to obtain $(\boldsymbol{P}^{\sf{B},{\star}},  \boldsymbol{P}^{\sf{S},{\star}}, \boldsymbol{\mu}^{\sf{B},{\star}}, \boldsymbol{\mu}^{\sf{S},{\star}})$.
		\STATE Update $ (\boldsymbol{P}^{{\sf{B}}, {(i)}}, \! \boldsymbol{P}^{{\sf{S}}, {(i)}}, \boldsymbol{\mu}^{{\sf{B}}, {(i)}}, \boldsymbol{\mu}^{{\sf{S}}, {(i)}}) \!\! := (\boldsymbol{P}^{\sf{B},{\star}},  \boldsymbol{P}^{\sf{S},{\star}}, \boldsymbol{\mu}^{{\sf{B}}, \star}, \boldsymbol{\mu}^{{\sf{S}}, \star})$.
        \STATE Set $i=i+1$.
		% \STATE 
        % \STATE Calculate $a_{n,k,s}^{(i)}, b_{n,k,s}^{(i)}$ and $\bp^{(i)}[t] = \exp(\bar{\bp}^{(i)}[t])$.
		\UNTIL Convergence
        % \STATE Calculate $d_k[t], \forall k$ based on \eqref{eq: data remain}.
        % \STATE Set $t=t+1$.
		% \ENDWHILE
        \STATE Recovery UA solutions $\boldsymbol{\alpha}$ and $\boldsymbol{\beta}$ by \eqref{eq: recover binary}.
        \STATE \textbf{Output:} The solution for problem $(\mathcal{P}_{0})$ $(\boldsymbol{\alpha}^{\star}, \boldsymbol{\beta}^{\star}, \boldsymbol{P}^{\sf{B},{\star}},  \boldsymbol{P}^{\sf{S},{\star}})$.
    \end{algorithmic}
    \normalsize 
\end{algorithm}

\subsection{Prediction-based Solution}

\begin{figure}
    \centering
    \includegraphics[width=85mm]{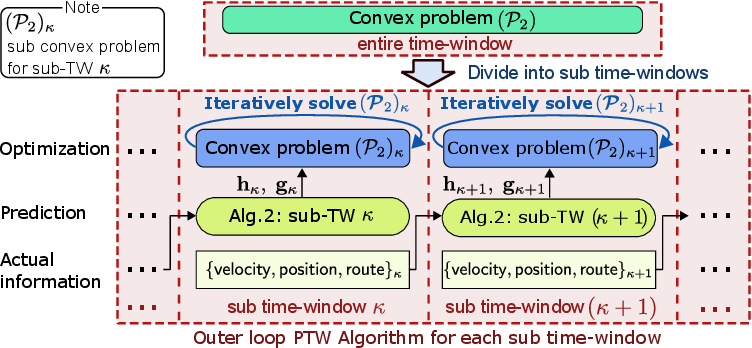}
    \vspace{-1mm}
    \captionsetup{font=small}
    \caption{Flowchart of the prediction-time-window algorithm}
    \label{fig:flowchart_predwindowAlg}
    % \vspace{-3mm}
\end{figure}
\begin{figure}
	\centering
	\includegraphics[width=70mm]{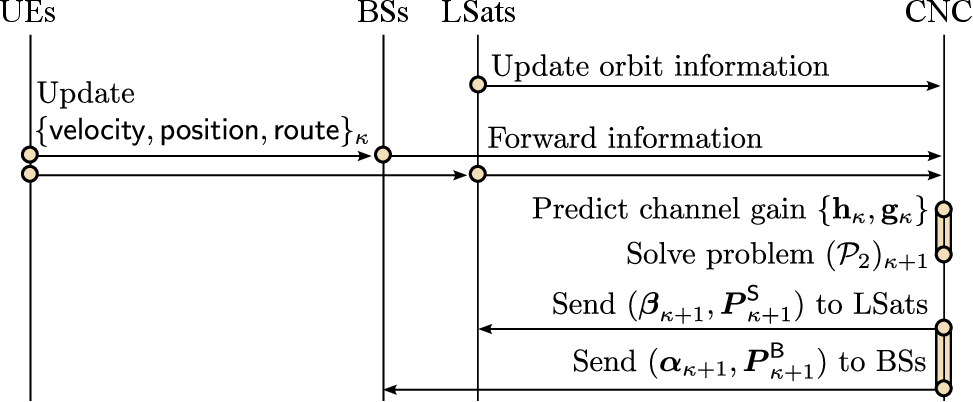}
	\vspace{-1mm}
	\captionsetup{font=small}
	\caption{Information exchange in the prediction-based algorithm}
	\label{fig:PredAlg_signaling}
	\vspace{-2mm}
\end{figure}
The implementation of the proposed FTW algorithm above requires the channel gain information of all TSs in the considered time window which is difficult to achieve in practice if the number of TSs is sufficiently large. Hence, in this sub-section, we further propose a prediction-based solution. The overall approach of this solution is described in Fig.~\ref{fig:flowchart_predwindowAlg}. Particularly, the entire TW is divided into multiple sub-TWs with the size of $N_{\sf{TS}}^{\sf{pred}}$, yielding $N_{\sf{TW}}^{\sf{pred}} = \lceil N_{\sf{TS}}/N_{\sf{TS}}^{\sf{pred}} \rceil$ sub-TWs. Subsequently, the channel gain information of each sub-TW is predicted based on the UE's information in the previous one. For convenience, let $\mathcal{T}_{\kappa}^{\sf{pred}} = \{ (\kappa - 1) N_{\sf{TS}}^{\sf{pred}}+1, \dots ,  \kappa  N_{\sf{TS}}^{\sf{pred}} \}$ be the set of TSs in sub-TW $\kappa$.

Supposing that the satellite orbit information, BS's position, and 3D map knowledge are available, one can further assume that the channel gain of LSat/BS-UE links can be identified by employing the RT mechanism for the given UE's position. 
However, to obtain the channel gain information in a sub-TW while avoiding the frequent update of UE's position at each TS, one necessitates predicting the UE position based on UE's information in the past. 
It is worth noting that UE is the vehicle, for the given UE route which can be obtained by the navigator application, e.g., Google Navigator, the UE position can be predicted as follows.
Regarding the UE mobility, it can be seen that the changing of the UE's velocity and trajectory in each sub-TW is negligible for a sufficiently small window size. Hence, let $v_{k,t}$ and $\hat{v}_{k,t}$ be the actual and predicted velocities of ${\sf{UE}}_{k}$ at TS $t$, ${\sf{UE}}_{k}$'s velocity in sub-TW $\kappa$ is predicted as
\vspace{-2mm}
\beq \label{eq: pred v}
    \hat{v}_{k,t} = \scaleobj{0.8}{\frac{1}{N_{\sf{TS}}^{\sf{pred}}} \sum\nolimits_{\forall t' \in \mathcal{T}_{\kappa - 1}^{\sf{pred}}}} v_{k,t'}, \quad \forall t \in \mathcal{T}_{\kappa}^{\sf{pred}}, \forall k.
\eeq
Subsequently, let $d_{k,t}$ be the distance from the starting point to the ${\sf{UE}}_{k}$'s position at TS $t$ along the given route, the predicted distance $\hat{d}_{k,t}$ in sub-TW $\kappa-$th can be identified as
\vspace{-2mm}
\beq \label{eq: pred d}
    \hat{d}_{k,t} = d_{k, t'} + T_{\sf{S}} \hat{v}_{k,t}, \quad \forall t \in \mathcal{T}_{\kappa}^{\sf{pred}}, 
\eeq
where $t' \! = \! (\kappa - 1) N_{\sf{TS}}^{\sf{pred}}$ is the TS before sub-TW $\kappa$. Based on the predicted UE position, the channel gain information can be identified by the RT mechanism. The channel gain prediction procedure is summarized as in Algorithm~\ref{alg: pred chan}.
Given the channel gain in sub-TW $\kappa$, the optimization problem in sub-TW $\kappa$ is revised from problem $(\mathcal{P}_{2})$ as
\vspace{-2mm}
\bieq {lll}\label{eq: SRCC Prob Pred}
(\mathcal{P}_{2})_{\kappa}: & \!
\max_{{ \boldsymbol{P}^{\sf{B}}_{\kappa}, \boldsymbol{P}^{\sf{S}}_{\kappa}, \boldsymbol{\lambda}^{\sf{B}}_{\kappa},  \boldsymbol{\lambda}^{\sf{S}}_{\kappa}, \boldsymbol{\mu}^{\sf{B}}_{\kappa}, \boldsymbol{\mu}^{\sf{S}}_{\kappa}, \boldsymbol{a}_{\kappa}, \boldsymbol{b}_{\kappa} }}  & \; \frac{\rho}{N_{\sf{TS}}^{\sf{pred}}} \tilde{\lambda}_{\kappa} - \frac{1 - \rho}{N_{\sf{TS}}^{\sf{pred}}} \tilde{f}^{{\sf{CC}},(i)}(\boldsymbol{P}^{\sf{B}}_{\kappa}, \boldsymbol{P}^{\sf{S}}_{\kappa}) \nonumber \\
&& \; \st \text{ constraints } (\tilde{C}1)_{\kappa} - (\tilde{C}12)_{\kappa}, \nonumber
\eieq
wherein the added low suffix $(\cdot)_{\kappa}$ in the constraint and variable notations indicates the adaptation of TS range which is limited within sub-TW $\kappa$. 
For simplification, we assume that sub-TW size $N_{\sf{TS}}^{\sf{pred}}$ is equal to integer times $N_{\sf{TS}}^{\sf{QoS}}$. 
To obtain the solution for problem $(\mathcal{P}_{0})$, channel gain prediction is executed and problem $(\mathcal{P}_{2})_{\kappa}$ is solved for each sub-TW. Particularly, at the end of sub-TW $(\kappa-1)$, channel gain for sub-TW $\kappa$ is predicted and problem $(\mathcal{P}_{2})_{\kappa}$ is solved. The proposed prediction-based algorithm, so-called prediction-based time-window (PTW) algorithm, is described in Algorithm~\ref{alg: OpzPower Pred}.
Its convergence is discussed in the following proposition.
\begin{proposition} \label{pro: PredAlg converg}
    Algorithm~\ref{alg: OpzPower Pred} is guaranteed for convergence to a local optimal solution of problem $(\mathcal{P}_{0})$.
    \vspace{-2mm}
\end{proposition}

\begin{IEEEproof} \label{app: proof PredAlg converg}
    To proof the convergence of Algorithm~\ref{alg: OpzPower Pred}, one needs to proof the independence among problems $(\mathcal{P}_{2})_{\kappa}$ of different sub-TWs and the convergence in solving each problem $(\mathcal{P}_{2})_{\kappa}$.
    Let $\Omega_{\kappa} = ( \boldsymbol{P}^{\sf{B}}_{\kappa}, \boldsymbol{P}^{\sf{S}}_{\kappa}, \boldsymbol{\lambda}^{\sf{B}}_{\kappa},  \boldsymbol{\lambda}^{\sf{S}}_{\kappa}, \boldsymbol{\mu}^{\sf{B}}_{\kappa}, \boldsymbol{\mu}^{\sf{S}}_{\kappa}, \boldsymbol{a}_{\kappa}, \boldsymbol{b}_{\kappa} ) $ be the variable tuple of problem $(\mathcal{P}_{2})_{\kappa}$. It can be seen that there is no coupling between variable tuples of problems $(\mathcal{P}_{2})_{\kappa}$ and $(\mathcal{P}_{2})_{\kappa'}, \forall \kappa \neq \kappa'$, i.e., $\Omega_{\kappa} \cap \Omega_{\kappa'} = \varnothing$. Therefore, problem $(\mathcal{P}_{2})_{\kappa}$ for each sub-TW $\kappa$ can be solved independently. In addition, problem $(\mathcal{P}_{2})_{\kappa}$ has the same structure with problem $(\mathcal{P}_{2})$. Hence, solving each problem $(\mathcal{P}_{2})_{\kappa}$ in steps $4$ of Algorithm~\ref{alg: OpzPower Pred} is guaranteed to converge, and the aggregated solution for all sub-TWs is the local optimal solution of problem $(\mathcal{P}_{0})$ which can be proved similarly to the proof of Proposition~\ref{pro: FullWAlg converg}.
\end{IEEEproof}

\begin{remark}
    The proposed FTW and PTW algorithms are designed based on the average joint max-sum-rate and min-CC-number problems $(\mathcal{P_{2}})$ and $(\mathcal{P}_{2})_{\kappa}$, however, they can easily adapt to the average joint max-sum-rate and min-CC-number objective with simple modifications. Particularly, let $\lambda_{t}^{\sf{min}}$ be the slack variable indicating the minimum UE rate at TS $t$. Indeed, by the following modifications:
    \begin{itemize}
        \item replace $\tilde{\lambda}$ by $\lambda_{t}^{\sf{min}}$ in the objective function,
        \item replace the rate requirement $\bar{R}_{k}$ by $\lambda_{t}^{\sf{min}}$ in $(\tilde{C}6)$ and $(\tilde{C}6)_{\kappa}$ of problems $(\mathcal{P}_{2})$ and $(\mathcal{P}_{2})_{\kappa}$, respectively,
    \end{itemize}
    the proposed FTW and PTW algorithms can be directly applied to solve average joint max-min-rate and min-CC-number problems.
    \vspace{-4mm}
\end{remark}

\vspace{-2mm}
\subsection{Prediction-based Algorithm Implementation}
\vspace{-1mm}
%\begin{figure}
%    \centering
%    \includegraphics[width=80mm]{Figures/PredAlg_signaling.eps}
% 	\vspace{-1mm}
%    \captionsetup{font=small}
%    \caption{Information exchange in the prediction-based algorithm}
%    \label{fig:PredAlg_signaling}
%    \vspace{-2mm}
%\end{figure}

To clarify the practical implementation, Fig.~\ref{fig:PredAlg_signaling} depicts the update and execution procedures of the prediction-based algorithm. In particular, at the end of each sub-TW $\kappa$, UEs update their velocity, position, and route via the connected BS/LSat while LSats update their orbit information to the CNC. Subsequently, the CNC executes Algorithm~\ref{alg: pred chan} to predict the channel gain for the next sub-TW $(\kappa+1)$. Based on which the CNC executes problem $(\mathcal{P}_{2})_{\kappa + 1}$ to find the solution for the next sub-TW $(\kappa+1)$. Since the update and execution procedures occur only once in each sub-TW, this approach reduces the signaling amount between nodes and the computation frequency at CNC. Additionally, depending on the stability of satellite orbits, LSats can update information less frequently instead of at every sub-TW which further reduces the data exchange frequency over the long propagation link between LSats and ground segments. Furthermore, the implementation analysis with the specific parameter setting is provided in the simulation results section.
\vspace{-2mm}

\begin{remark}
    Thanks to updating the UE route at each sub-TW, the PTW algorithm can adapt to the changes of the UE route in practice which avoids the outdated route information provided at the beginning time.
    \vspace{-1mm}
\end{remark}

\begin{remark}
    In practical implementation, the resource allocation task should be completed within a finite time horizon, which may vary depending on the practical systems. For simplification, we assume that the computation capacity of the CNC is adequate and the time horizon for the resource allocation task is extensive enough. 
\end{remark}

\begin{algorithm}[!t]
\footnotesize
    \begin{algorithmic}[1]
 \captionsetup{font=small}
		\protect\caption{\textsc{Channel Gain Prediction}}
		\label{alg: pred chan}
		\STATE \textbf{Input:} Satellite orbit, BS's positions, 3D map data, UE's routes, sub-TW index $\kappa$, and UE's velocity and position in the previous sub-TWs.
        % \STATE Set $t=1$.\\
        % \WHILE {$t \leq N_T$ or $d_k[t]=0, \forall k$}
        \STATE Predict $\hat{v}_{k,t}$ and $\hat{d}_{k,t}$ of UEs in sub-TW $\kappa$ by \eqref{eq: pred v} and \eqref{eq: pred d}.
        \STATE Identify UE positions from $\hat{d}_{k,t}$ with respect to given UE's routes.
        \STATE Identify channel gain from UE position by RT mechanism.
        \STATE \textbf{Output:} The channel gain information of UEs in sub-TW $\kappa$.
    \end{algorithmic}
    \normalsize 
\end{algorithm}

\begin{algorithm}[!t]
\footnotesize
	\begin{algorithmic}[1]
        \captionsetup{font=small}
		\protect\caption{\textsc{PTW Algorithm for Solving Problem $(\mathcal{P}_{0})$}}
		\label{alg: OpzPower Pred}
		% \long\def\algorithmicrequire{\textbf{Phase 1:}}
		% \REQUIRE
        \STATE Set $\kappa=1$.
        \WHILE {$\kappa \leq N_{\sf{TW}}^{\sf{pred}}$}
        \STATE Predict the channel gain in sub-TW $\kappa$ by Algorithm~\ref{alg: pred chan}.
        \STATE Solve problem $(\mathcal{P}_{2})_{\kappa}$ by Algorithm~\ref{alg: OpzPower FullTS} with adaptation for sub-TW $\kappa$.
        \STATE Set $\kappa=\kappa+1$.
		\ENDWHILE
        \STATE Recovery UA solutions $\boldsymbol{\alpha}$ and $\boldsymbol{\beta}$ by \eqref{eq: recover binary}.
        \STATE \textbf{Output:} The solution $(\boldsymbol{\alpha}^{\star}, \boldsymbol{\beta}^{\star}, \boldsymbol{P}^{\sf{B},{\star}},  \boldsymbol{P}^{\sf{S},{\star}})$.
    \end{algorithmic}
    \normalsize 
\end{algorithm}

\section{Benchmarks and Complexity Analysis} \label{sec: benchmark, complexity}
\subsection{Greedy Algorithm}
In this benchmark algorithm, the UA is selected based on the channel gain while the transmit power of each BS/LSat is allocated uniformly.
For BS-UE connection, at each TS, the $\sf{BS}_{n}-\sf{UE}_{k}$ link is chosen based on the descending channel gain order if: 
1) $\sf{UE}_{k}$ is still not connected to any BSs to ensure $(C1)$
and 2) the total number of connected and selected UEs at $\sf{BS}_{n}$ is still less than $\Psi_{n}^{\sf{B}}$ to ensure $(C2)$. The same procedure is conducted for the LSat-SUE connection such that constraints $(C3)$ and $(C4)$ are satisfied. Since only the channel gain-based UA is considered, the QoS constraint $(C6)$ is not guaranteed. The Greedy Algorithm is described in Algorithm~\ref{alg: greedy}. For convenience, the MATLAB notation is used to perform the matrix index.

\begin{algorithm}[t]
\footnotesize
\begin{algorithmic}[1]
     \captionsetup{font=small}
    \protect\caption{\textsc{Greedy Algorithm}}
    \label{alg: greedy}
    \STATE \textbf{Input:} Channel gain $ \mathbf{h}=[h_{n,k,t}]_{\forall (n,k,t)} $ and $ \mathbf{g} = [g_{m,k,t}]_{\forall (m,k,t)}$, and parameters $\{ \psi_{n}^{\sf{B}}, \psi_{m}^{\sf{S}}, \eta_{n,t}^{{\sf{B}}}, \eta_{m,t}^{{\sf{S}}} \}_{\forall (m,n,t)}$.
    \STATE \textbf{Initialize:} Zero matrices $\boldsymbol{\alpha}$ and $\boldsymbol{\beta}$
    \FOR{Each TS $t$}
    \WHILE{$\mathbf{h}(:,:,t) \neq \mathbf{0}$}
    \STATE Find index $(\hat{n},\hat{k})$ satisfying $\mathbf{h}(\hat{n},\hat{k},t) = {\sf{max}}( \mathbf{h}(:,:,t))$.
    \STATE \textbf{if} {$\scaleobj{0.8}{\sum_{\forall k}} \alpha_{\hat{n},k,t}  \leq \psi_{\hat{n}}^{\sf{B}} - \eta_{\hat{n},t}^{{\sf{B}}}$}
    \STATE \textbf{then} Update $\alpha_{\hat{n},\hat{k},t} = 1$
         and $\mathbf{h}(:,\hat{k},t) = \mathbf{0}$
    % \ELSE
        \STATE \textbf{else} Update $\mathbf{h}(\hat{n},:,t) := \mathbf{0}$
    % \ENDIF
    \ENDWHILE
    \WHILE{$\mathbf{g}(:,:,t) \neq \mathbf{0}$}
    \STATE Find index $(\hat{m},\hat{k})$ satisfying $\mathbf{g}(\hat{m},\hat{k},t) = {\sf{max}}( \mathbf{g}(:,:,t))$.
    \STATE \textbf{if} {$\scaleobj{0.8}{\sum_{\forall k}} \beta_{\hat{m},k,t}  \leq \psi_{\hat{m}}^{\sf{S}} - \eta_{\hat{m},t}^{{\sf{S}}}$}
    \STATE \textbf{then} Update $\beta_{\hat{m},\hat{k},t} = 1$
         and $\mathbf{g}(:,\hat{k},t) = \mathbf{0}$
    % \ELSE
        \STATE \textbf{else} Update $\mathbf{g}(\hat{m},:,t) := \mathbf{0}$
    % \ENDIF
    \ENDWHILE
    \ENDFOR
    \STATE \textbf{Output:} UA solution $\boldsymbol{\alpha}, \boldsymbol{\beta}$.
    % \vspace{-2mm}
\end{algorithmic} 
\normalsize
% \vspace{-2mm}
\end{algorithm}

\subsection{Other Benchmark Algorithm}
For comparison purposes, we introduce a benchmark algorithm based on our previous work \cite{Hung_VTC24}. In \cite{Hung_VTC24}, we studied a similar system model to that in this work wherein the UA is optimized with a fixed transmit power. However, in \cite{Hung_VTC24} the objective is only SR maximization. To adapt the algorithm proposed in \cite{Hung_VTC24} to this work, we aim to modify this algorithm as follows. To consider the multi-objective in problem $(\mathcal{P}_{0})$, the CC minimization objective $f^{\sf{CC}}(\boldsymbol{\alpha}, \boldsymbol{\beta})$ is added, wherein multi-objective factors $\rho$ and $(1 - \rho)$ are added into SR and CC objectives. Subsequently, proposition~\ref{pro: up bound fCC} is applied to $f^{\sf{CC}}(\boldsymbol{\alpha}, \boldsymbol{\beta})$, where components $|\alpha_{n,k,t} - \alpha_{n,k,t-1}|$ and $| \beta_{m,k,t} - \beta_{m,k,t-1}|$ are convexified similar to the procedure applied to $| \Vert \boldsymbol{P}_{n,k,t}^{\sf{B}} \Vert_{0} - \Vert \boldsymbol{P}_{n,k,t-1}^{\sf{B}} \Vert_{0}|$ and $| \Vert \boldsymbol{P}_{m,k,t}^{\sf{S}} \Vert_{0} - \Vert \boldsymbol{P}_{m,k,t-1}^{\sf{S}} \Vert_{0}|$, respectively. Additionally, the transmit power at each BS/LSat is fixed and uniformly allocated. This modified algorithm is named the full-time-window UA (FWUA) algorithm.

\subsection{Complexity Analysis}
This section analyzes the complexity of the proposed algorithms and benchmark algorithms regarding big-O notation. One notes that FTW, PTW, and FWUA algorithms involve solving convex problems. Assuming that the interior method is used to solve the convex problems, the complexity in terms of big-O is $\mathcal{O}(c^{1/2} (c+v) v^2)$ where $c$ and $v$ are the numbers of inequality constraints and variables, respectively \cite{complexity}. 

\subsubsection{FTW Algorithm}
As described in Algorithm~\ref{alg: OpzPower FullTS}, the FTW Algorithm includes one loop in \textbf{Steps 2-6}, wherein problem $(\mathcal{P}_{2})$ is solved iteratively. Since problem $(\mathcal{P}_{2})$ consists of $v_{\sf{FTW}} = 3NK N_{\sf{TS}} + 3MK N_{\sf{TS}} + N N_{\sf{TS}} + M  N_{\sf{TS}}$ variables and $c_{\sf{FTW}} = 4NK N_{\sf{TS}} + 4MK N_{\sf{TS}} + 2 N N_{\sf{TS}} + 2 M N_{\sf{TS}} + 5K N_{\sf{TS}} + K N_{\sf{TS}} / N_{\sf{TS}}^{\sf{QoS}}$ inequality constraints, the complexity of the FTW Algorithm can be expressed as
\beq
    X_{\sf{FTW}} = \mathcal{O}(N_{\sf{iter}} c_{\sf{FTW}}^{1/2} (c_{\sf{FTW}} + v_{\sf{FTW}}) v_{\sf{FTW}}^3 ),
\eeq
wherein $N_{\sf{iter}}$ is the required number iteration for convergence.

\subsubsection{PTW Algorithm}
Regarding Algorithm~\ref{alg: OpzPower Pred}, the PTW Algorithm consists of two loops: the outer loop corresponds to the number of sub-TWs $N_{\sf{TW}}^{\sf{pred}}$  and the inner loop relates to solving problem $(\mathcal{P}_{2})_{\kappa}$ iteratively. Since problem $(\mathcal{P}_{2})_{\kappa}$ consists of $v_{\sf{PTW}} = 3NK N_{\sf{TS}}^{\sf{pred}} + 3MK N_{\sf{TS}}^{\sf{pred}} + N N_{\sf{TS}}^{\sf{pred}} + M  N_{\sf{TS}}^{\sf{pred}}$ variables and $c_{\sf{PTW}} = 4NK N_{\sf{TS}}^{\sf{pred}} + 4MK N_{\sf{TS}}^{\sf{pred}} + 2 N N_{\sf{TS}}^{\sf{pred}} + 2 M N_{\sf{TS}}^{\sf{pred}} + 5K N_{\sf{TS}}^{\sf{pred}} + K N_{\sf{TS}}^{\sf{pred}} / N_{\sf{TS}}^{\sf{QoS}} $ inequality constraints, the complexity of the PTW Algorithm can be expressed as
\beq
    X_{\sf{PTW}} = \mathcal{O}( N_{\sf{iter}} N_{\sf{TW}}^{\sf{pred}} c_{\sf{PTW}}^{1/2} (c_{\sf{PTW}} + v_{\sf{PTW}}) v_{\sf{PTW}}^3 ).
\eeq

\subsubsection{Greedy Algorithm}
Regarding Algorithm~\ref{alg: greedy}, the Greedy Algorithm consists of one outer-loop with $N_{\sf{TS}}$ loops and two inner-loops for BS-UE and LSat-UA, respectively. In the first inner-loop in \textbf{Step 4-8}, one needs to execute $K$ times until $\mathbf{h}(:,:,t)=\mathbf{0}$ in worst case, wherein finding and updating in \textbf{Step 5-8} demand a complexity of $\mathcal{O}(NK + N)$ for each time. The computation for each time in the second inner loop is identified similarly. In summary, the complexity of the Greedy Algorithm can be given as 
\beq
    X_{\sf{Gre.Alg.}} = \mathcal{O}(N_{\sf{TS}} ( NK^2 + NK + MK^2 + MK )).
\eeq

\subsubsection{FWUA Algorithm}
In the FWUA Algorithm, a convex problem for UA is solved iteratively until convergence. The convex problem associated with the FTWUA Algorithm consists of $v_{\sf{FWUA}} = 3NK N_{\sf{TS}} + 3MK N_{\sf{TS}} + N N_{\sf{TS}} + M  N_{\sf{TS}}$ variables and $c_{\sf{FWUA}} = 5NK N_{\sf{TS}} + 5MK N_{\sf{TS}} + N N_{\sf{TS}} + M N_{\sf{TS}} + 5K N_{\sf{TS}} + K N_{\sf{TS}} / N_{\sf{TS}}^{\sf{QoS}}$ inequality constraints. Therefore, the complexity of the FTWUA Algorithm can be expressed as
\beq
    X_{\sf{FWUA}} = \mathcal{O}(N_{\sf{iter}} c_{\sf{FWUA}}^{1/2} (c_{\sf{FWUA}} + v_{\sf{FWUA}}) v_{\sf{FWUA}}^3 ).
\eeq

\begin{figure}
    \centering
    \includegraphics[width=80mm]{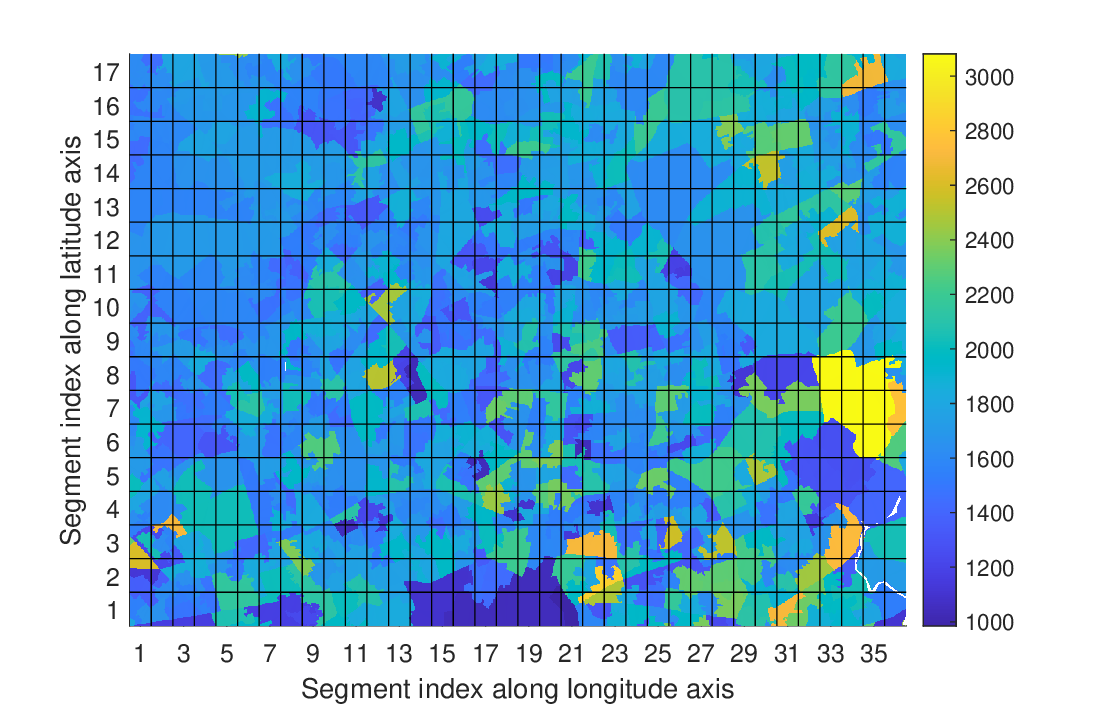}
        \vspace{-1mm}
    \captionsetup{font=small}
    \caption{ Population distribution in the entire examined area, the total population in each colored area is indicated by its color.}
    \label{fig:population_map}
\end{figure}

\begin{figure}
    \centering
    \includegraphics[width=80mm]{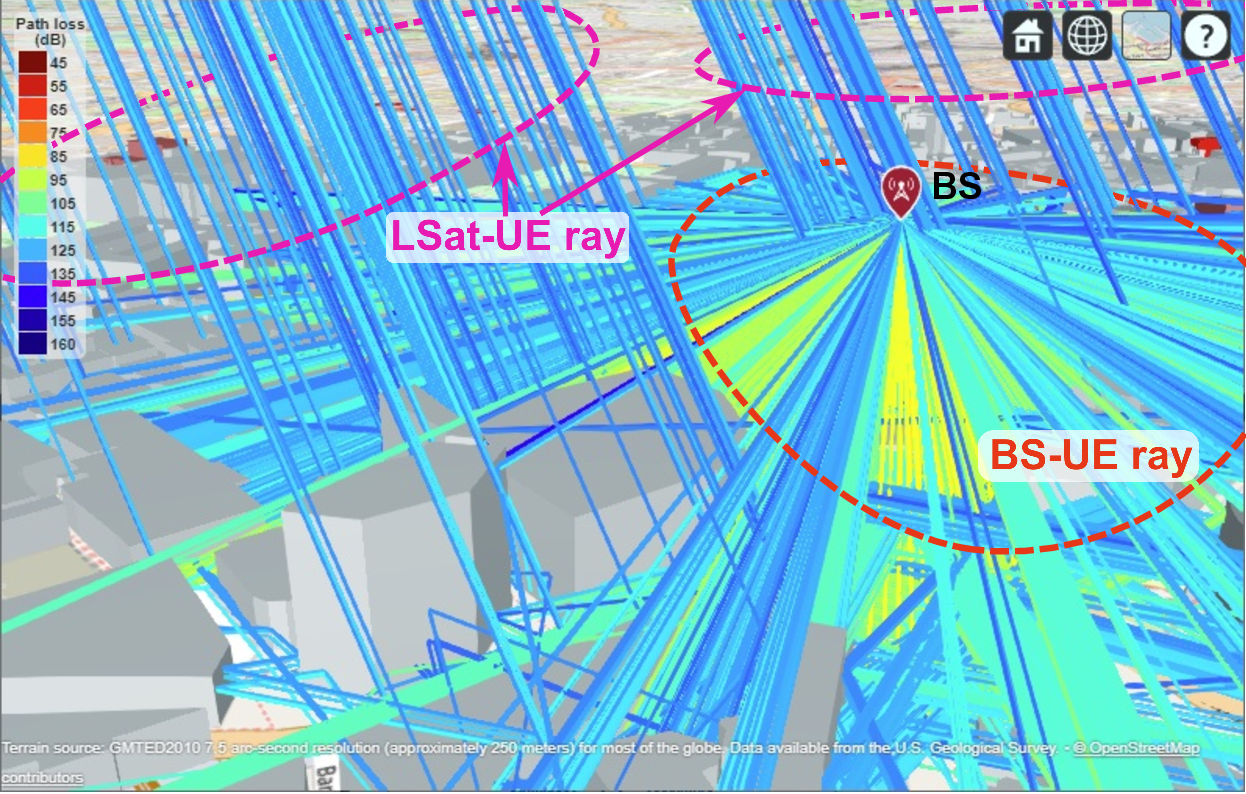}
    \vspace{-1mm}
    \captionsetup{font=small}
    \caption{Raytracing result of BS/LSat-UE links.}
    \label{fig:RT_result}
\end{figure}

\section{Numerical Results} \label{sec: results}

\subsection{Simulation Setup}
The simulation is conducted in urban environments on an area in London city limited by latitude and longitude ranges of $[51.5115 ^\circ \text{N}, 51.5965^\circ \text{N}]$ and $[0.1772^\circ \text{W} , 0.0028^\circ \text{E}]$, respectively.
Subsequently, the map is divided into a grid of segments where the size of each segment is $(0.005^\circ \times 0.005^\circ)$ in latitude and longitude, respectively. To evaluate the link budget across map locations, the receiver grid size in each map segment is $(100 \times 100)$ points. It is worth noting that UEs are the vehicles, thus receiver points are deployed outdoors. Regarding the transmitters, one assumes that one BS is deployed in each map segment at the highest building. 
In addition, one LEO constellation at $500$ km is considered. To reduce the shadow area caused by high-buildings when LSats at low elevation angle and improve the coverage, one assumes that there are $2$ LSats which can be used in the FoV at each instance time. 
Regarding UE mobility, the UE's routes are obtained by the navigator data in the Google Map application.
% Regarding the UE mobility, the UE's routes are obtained by the navigator data in the Google Map application as in Fig~\ref{fig:ggmap_plt}.
Additionally, the number of TUEs at ${\sf{BS}}_{n}$ $\eta_{n,t}^{\sf{B}}$ and SUEs at ${\sf{LSat}}_{m}$ $\eta_{m,t}^{\sf{S}}$ are assumed to be Poisson random variables with means $\bar{\eta}_{n}^{\sf{B}}$ and $\bar{\eta}_{m}^{\sf{S}}$. Furthermore, to consider the heterogeneity of UE number in the considered area, the means number of TUEs at BSs $\bar{\eta}_{n}^{\sf{B}}$ are set proportionally to the population distribution in the area which is extracted from London's age data set \cite{london_population} and depicted in Fig.~\ref{fig:population_map}. Particularly, the population in each map segment is computed based on the dataset, then $\bar{\eta}_{n}^{\sf B}$ for each map segment is scaled proportionally to its population so that $\max_{n}\{ \bar{\eta}_{n}^{\sf{B}}\}  = \bar{\eta}_{\max}^{\sf B}$.
The key simulation parameters are described in Table.~\ref{tab: parameter} wherein parameter set 1 for co-existence study in 3GPP TR 38-863 is utilized. In practical systems, if the power flux density limit is explicit, the maximum LSat transmit power could be adjusted appropriately.
The simulations are carried out on a computer equipped with an \textit{Intel Xeon E5 CPU @ 2.4~GHz} and \textit{128~GB RAM}, wherein the iterative algorithms are implemented by using the MATLAB programming environment, leveraging the CVX modeling framework and MOSEK solver.
For an intuitive view of complicated environments in urban areas, we first investigate the link budget assessment in terms of the BS/LSat-UE link's CINR with practical data. Subsequently, the numerical results are carried out to evaluate the performance in terms of SR and CC number optimization of the considered algorithms.

% \begin{figure*}[h!]
% \begin{subfigure}{0.35\textwidth}
%     \centering
%     \includegraphics[width=63mm, height=45mm]{Figures/CINR_BSUE_heatmap.png}
%     \caption{CINR BS-UE links.}
%     \label{fig:CINR_BSUE_heatmap}
% \end{subfigure}
% \begin{subfigure}{0.35\textwidth}
%     \centering
%     \includegraphics[width=63mm, height=45mm]{Figures/CINR_SatUE_heatmap.png}
%     \caption{CINR LSat-UE links.}
%     \label{fig:CINR_SatUE_heatmap}
% \end{subfigure}
% \begin{subfigure}{0.25\textwidth}
%     \centering
%     \vspace{15mm}
%     \includegraphics[height=35mm]{Figures/ggmap_plt.eps}
%     \caption{UE's route in GoogleMap.}
%     \vspace{5mm}
%     \label{fig:ggmap_plt}
% \end{subfigure}
% \vspace{-1mm}
% \captionsetup{font=small}
% \caption{CINR heatmap in dB scale and routes of $\sf{UE}_{1}$, $\sf{UE}_{7}$, and $\sf{UE}_{12}$.}
%      \label{fig:CINR_Heatmap_UE_trajec}
%      \vspace{-2mm}
% \end{figure*}

\begin{figure}
    \centering
    \includegraphics[width=85mm]{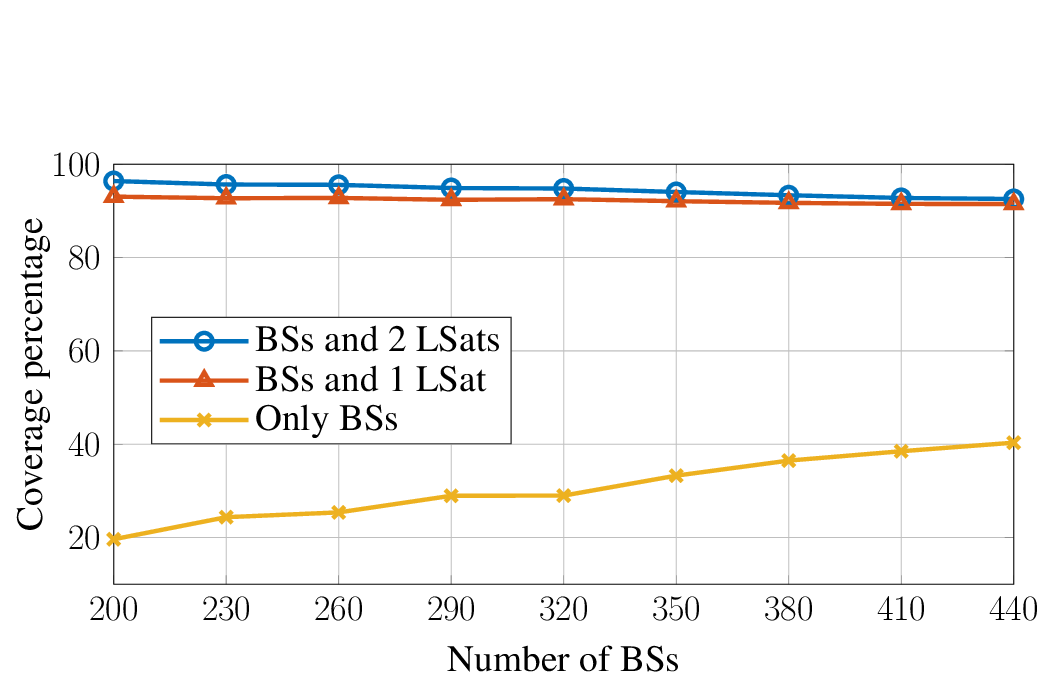}
    \vspace{-2mm}
    \captionsetup{font=small}
    \caption{The coverage percentage (CINR $
    \geq 3$ dB) versus the number of deployed BSs.}
    \label{fig:Cover_NumBSs}
\end{figure}

\begin{figure*}
\begin{subfigure}{0.65\textwidth}
\begin{subfigure}{1\textwidth}
    \centering
    \includegraphics[width=110mm]{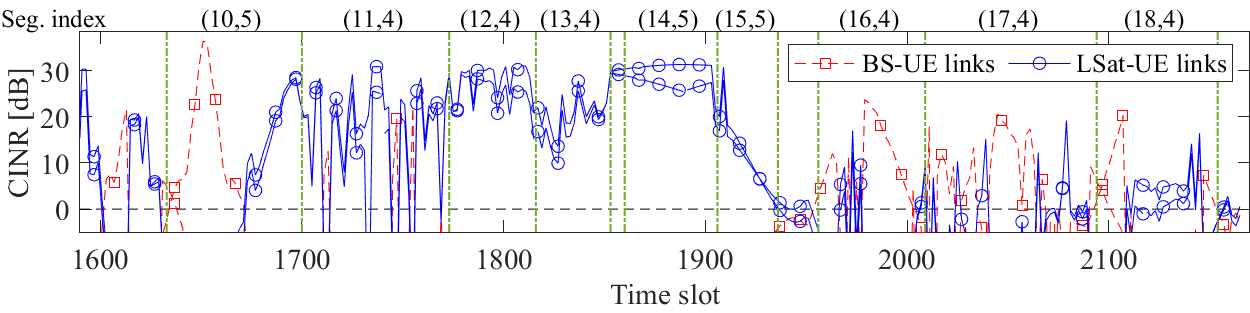}
    % \vspace{-6mm}
    % \caption{CINR of BS/LSat-$\sf{UE}_{1}$ links over TSs.}
    % \label{fig:CINR_time_UE1}
\end{subfigure}
\begin{subfigure}{1\textwidth}
    \centering
    \includegraphics[width=110mm]{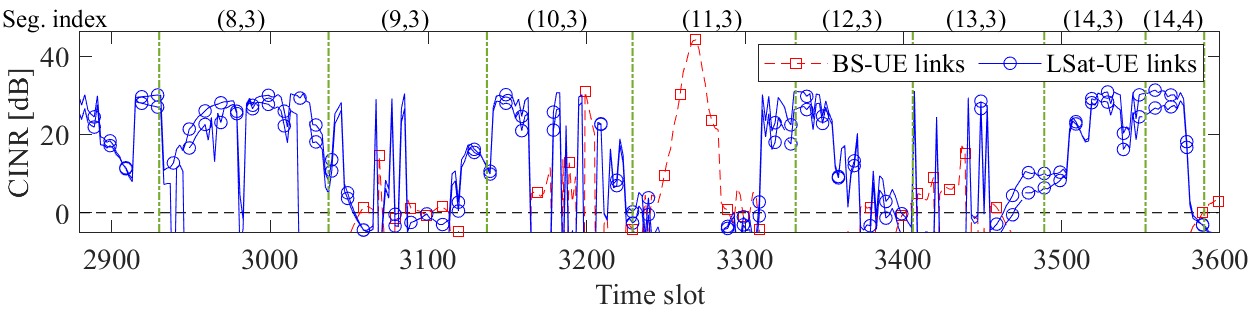}
    % \vspace{-6mm}
    \caption{CINR of BS/LSat-$\sf{UE}_{1}$ (above) and  BS/LSat-$\sf{UE}_{7}$ (below) links over TSs.}
    \label{fig:CINR_time_UE7}
\end{subfigure}
\end{subfigure}
\begin{subfigure}{0.35\textwidth}
\begin{subfigure}{1\textwidth}
    \centering
    \includegraphics[width=62mm]{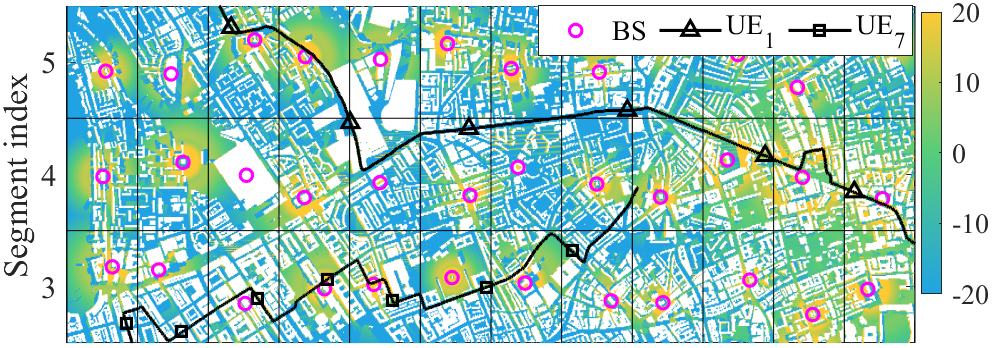}
    \caption{CINR of BS-UE links.}
    \label{fig:CINR_BSUE_heatmap_sub}
\end{subfigure}
\begin{subfigure}{1\textwidth}
    \centering
    \includegraphics[width=62mm]{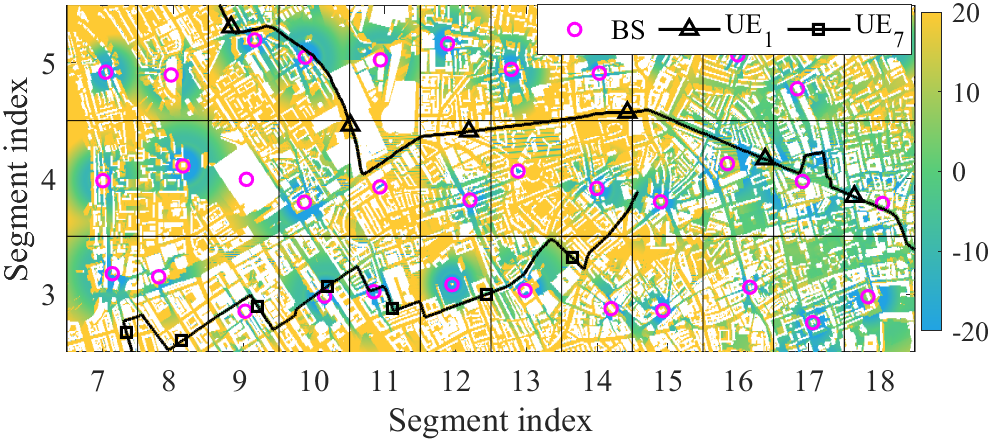}
    \caption{CINR of LSat-UE links.}
    \label{fig:CINR_SatUE_heatmap_sub}
\end{subfigure}
\end{subfigure}
\vspace{-5mm}
\captionsetup{font=small}
\caption{CINR of BS/LSat-$\sf{UE}_{1}$ and BS/LSat-$\sf{UE}_{7}$  links over TSs corresponding to the CINR heatmap in a small area.}
     \label{fig:CINR_time_UE_trajec}
     \vspace{-3mm}
\end{figure*}

\begin{table}[!t]
\captionsetup{font=footnotesize}
	\caption{Simulation Parameters}
	\label{tab: parameter}
	\centering
		\scalebox{0.95}{
	\begin{tabular}{l|l}
		\hline
		Parameter & Value \\
		\hline\hline
        Number of TSs, $N_{\sf{TS}}$            & $3600$ TSs\\
        Sub-TW size, $N_{\sf{TS}}^{\sf{pred}}$            & $90$ TSs\\
        TS duration, $T_{\sf{S}}$                         & $0.5$ s \\
        Operation frequency, $f_c$              & $3.4$ GHz \\
        System bandwidth, $B$   & $20$ MHz \\
        Latitude limitation of area             & $[51.5115 ^\circ \text{N}, 51.5965^\circ \text{N}]$   \\
        Longitude limitation of area            & $[0.1772^\circ \text{W} , 0.0028^\circ \text{E}]$   \\
        Size of one map segment                   & $0.005^\circ \times 0.005^\circ$ \\
        Receiver point grid size in one segment & $100 \times 100$ points      \\
        Number of BSs                           & $467$ \\
        BS antenna parameters                   & \cite{3gpp.38.863} \\
        LEO constellation                  & $500$ km \\
        LEO inclination angle   & $53^\circ$ \\
        Minimum elevation angle in FoV $\tilde{\theta}$ & $60^\circ$ \\
        Number of LSats at each time  & $2$ \\
        LSat antenna aperture, $a$    & $1$ m \cite{3gpp.38.821, 3gpp.38.863} \\
        Maximum LSat antenna gain, $G_{\sf{Sat}}^{\sf{max}}$   & $30$ dBi \cite{3gpp.38.821, 3gpp.38.863} \\
        Number of UEs           &   $12$ \\
        UE antenna height                         & $1$ m   \\
        Maximum UE antenna gain         & $12.8$ dBi \\
        BS's power budget, $P^{\sf{B,max}}_{n}$    &   $42$ dBm \\
        LSat's power budget, $P^{\sf{S,max}}_{m}$    &   $14$ dBW \\
        UE receiver noise figure, $G_f$ & $1.2$ dB \\
        UE antenna temperature, $T_a$  & $150$ K \\
        Penetration loss of IRR glass, $L_{\sf{IRRglass}}$ & $23 + 0.3f$[GHz] \cite{3gpp.38.901}   \\
        Penetration loss of concrete, $L_{\sf{concrete}}$ & $5 + 4f$[GHz] \cite{3gpp.38.901}   \\
        Maximum number of served UEs at BSs, $\psi_{n}^{\sf{B}}$  & $20$ \\
        Maximum number of served UEs at LSats, $\psi_{m}^{\sf{S}}$  & $100$ \\
        Maximum mean TUE number at BSs, $\bar{\eta}_{\max}^{\sf{B}} $ & $18$ \\
        Mean SUE number at LSats, $\bar{\eta}_{m}^{\sf{S}}$ & $95$ \\
        Multi-objective factor, $\rho$ & $0.9$  \\
        Smooth parameter of the approx. function, $\zeta$ & $10$ \\
		\hline		   				
	\end{tabular}}
	%	\vspace{-0.5pt}
\end{table}

\subsection{Link Budget Simulation Results} \label{sec:CINR results}
This section carries out the simulation to assess the link budget in terms of the CINR of BS/LSat-UE links in urban environments. The CINR of links can be calculated as described in \cite{Guidotti_ICC20_NTNlink}. Additionally, based on the outcomes of RT simulation, i.e., propagation loss, phase delay, and AoA and AoD of rays, the equivalent PL of each link is computed by \eqref{eq: PL model}. 
To be more intuitive, the RT simulation of the BS-UE and LSat-UE links is shown in Fig.~\ref{fig:RT_result}, where the elevation angle at the receiver points to LSat is $60^\circ$. This shows the impact of building and the complicated propagation of rays from BS/LSat to the receiver points in urban environments, which is difficult to determine mathematically.

To evaluate the effectiveness of ISTNs in improving coverage, Fig.~\ref{fig:Cover_NumBSs} illustrates the percentage of coverage area with $\text{CINR}\geq 3$ dB of different deployment options versus the number of BSs. Obviously, increasing the number of BSs improves coverage. However, based on the trend, one may require to deploy a large number of BSs to achieve more than $90$\% coverage. In contrast, using integrated BSs with $1$ or $2$ LSats can significantly increase the coverage percentage, i.e., more than $90$\% of the considered area. This also demonstrates the cost-effectiveness of ISTNs in improving coverage and facilitating seamless connectivity.

Fig.~\ref{fig:CINR_time_UE_trajec} shows the CINR heatmap of BS/LSat-UE links and the time-varying CINR of two UEs corresponding to their route in a considered area. Regarding the CINR heatmap, one can see the heterogeneity of both BS-UE and LSat-UE links. In particular, the CINR of BS-UE links is improved in areas with dense buildings and dense BS deployment. This is due to the blockage of the high and dense building to the LSat-UE links while the BS-UE links against the impact of the building by dense deployment of BSs as depicted in Fig.~\ref{fig:RT_result}. Additionally, the CINR of LSat-UE links increases in the area with fewer buildings.
Interestingly, despite the significant propagation loss due to the long distance of the LSat-UE links compared to BS-UE ones, its CINR is higher than that of the BS-UE link in certain areas. This phenomenon is due to the impact of the antenna gain pattern at the UE which is equipped with the patch antenna. In particular, the elevation angle at the UE towards the LSat can be seen as constant across the map due to the long UE-LSat distance as in Fig.~\ref{fig:RT_result}. 
This elevation angle is in a range of about $60^\circ - 90^\circ$, which results in an antenna gain at UE of about $7.3 - 12.8$ dBi as shown in Fig.~\ref{fig:2x2PatchPattern}. However, except for locations near BSs, the BS-UE links are transmitted almost in the horizontal plane with a low elevation angle, i.e., about $20^\circ - 30^\circ$, leading to the low antenna gain at UE, i.e., about $-26.3$ dBi to $-12.8$ dBi. For instance, at locations wherein the elevation angles of LSat-UE and BS-UE links are $60^\circ$ and $30^\circ$, the gap in antenna gain between two links is about $20.1$ dBi. Therefore, in areas where there exists a significant antenna gain disparity, the antenna gain can compensate the distance gap between the LSat-UE and BS-UE links, allowing the LSat-UE links to outperform the BS-UE links in terms of CINR. 
Additionally, regarding time-varying CINR, one can see the correlation between the UE routes on the CINR heatmap and their CINR over TSs. This change in both BS/LSat-UE links is due to the combination of UE mobility, LSat movement, and building influence. Depending on the UE location, the quality of BS/LSat-UE links is different, especially the LSat-UE links which are time-dependence due to LSat movement. In particular, the CINR can be changed with different positive magnitudes in a range of about $5-30$ dB or even be negative, i.e., the link is blocked. This emphasizes the necessity of PA and UA optimization as formulated in problem $(\mathcal{P}_{0})$.

\subsection{Numerical Results}
\begin{figure}
    \centering
    \includegraphics[height=45mm, width=80mm]{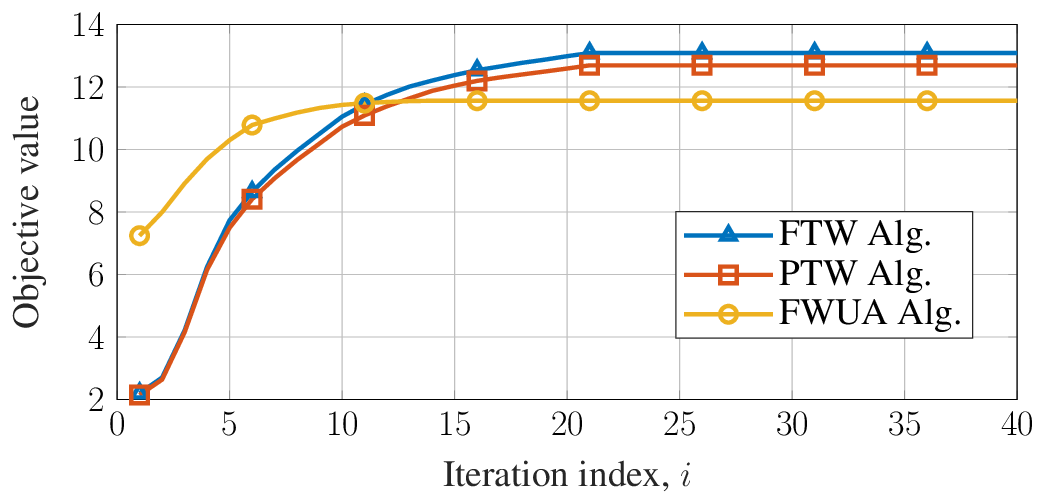}
    \vspace{-2mm}
    \captionsetup{font=small}
    \caption{Convergence of the objective function.}
    \label{fig:Obj_convergence}
\end{figure}

This section discusses the numerical results of the proposed and benchmark algorithms in solving the SR-CC problem.
Fig.~\ref{fig:Obj_convergence} shows the convergence of the objective function which consists of the average SR and the CC number of the FTW, PTW, and FTUA algorithms versus iterations. All considered algorithms require only a few tens iterations for convergence. Due to the optimization of only UA, the FWUA algorithm requires the smallest number of iterations for convergence among three, about $14$ iterations. Additionally, the proposed FTW and PTW algorithms need a slightly higher number of iterations, i.e., about $23$ and $24$ iterations, respectively. In practical implementation, if the running time of the proposed algorithms is too long, one can stop the execution before convergence to obtain a good enough solution.

\begin{figure}
    \centering
    \includegraphics[width=80mm]{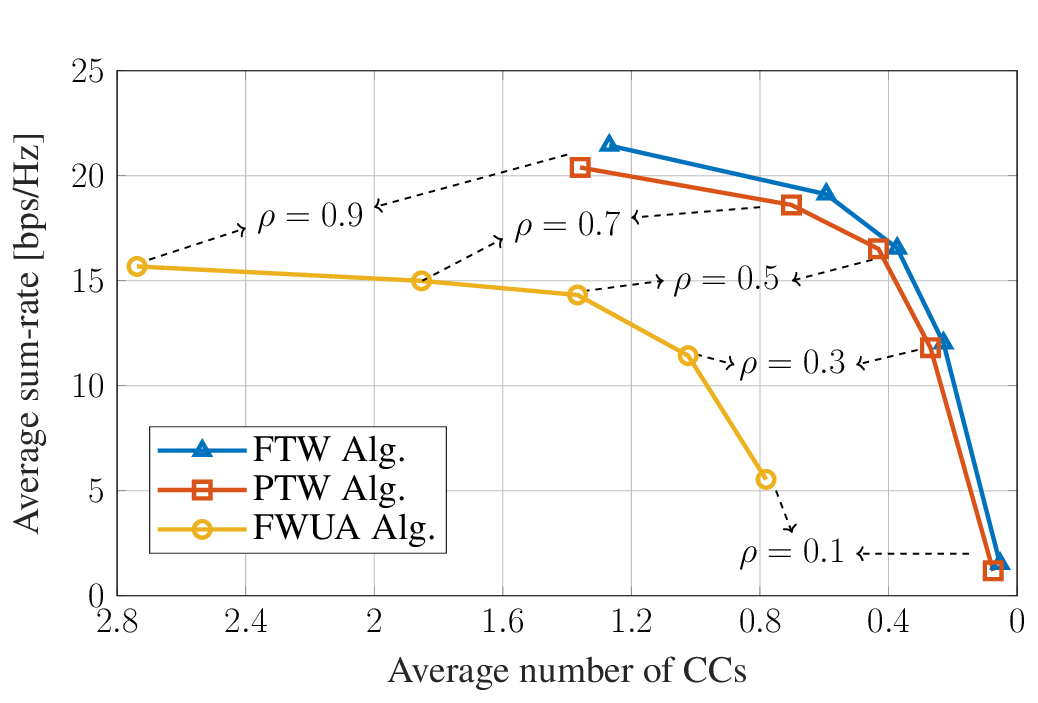}
    \vspace{-2mm}
\captionsetup{font=small}
    \caption{Pareto front of the average SR and CC number vs multi-objective factor $\rho$.}
    \label{fig:SR_HO_rho}
\end{figure}

Fig.~\ref{fig:SR_HO_rho} depicts the Pareto curves of the average SR and CC number per TS of the considered algorithms versus multi-objective factor $\rho$. 
One can see that the Pareto curves of the two proposed algorithms are close and outperform that of the FWUA algorithm in terms of the SR and the CC number. Particularly, at $\rho=0.9$, $\rho=0.7$, and $\rho=0.5$, both SR and the CC number results provided by the two proposed algorithms are better than those by the FWUA algorithm. At $\rho=0.3$ and $\rho=0.1$, the FWUA algorithm can provide a better SR compared to the two proposed algorithms, however, the CC number outcome is worse.
Obviously, a higher $\rho$ increases the priority of SR maximization and decreases that of CC number minimization. However, for the FTW and PTW algorithms, the Pareto curve increases quickly from $\rho=0.1$ to $\rho=0.7$, i.e., SR rapidly raises and the CC number slowly increases, whereas it slowly increases from $\rho=0.7$ to $\rho=0.9$. Particularly, SR increases about $10.6$, $5.9$, and $2.6$ bps/Hz and the average CC number increases about $0.15$, $0.23$, and $0.35$ when increasing $\rho=0.1 \rightarrow 0.3$, $\rho=0.3 \rightarrow 0.5$, and $\rho=0.5 \rightarrow 0.7$, whereas the average SR and the CC number increase about $2.27$ bps/Hz and $0.6$ when increasing $\rho=0.7 \rightarrow 0.9$, respectively. Based on this observation, $\rho=0.7$ is selected to balance between the SR and CC number.
% \begin{figure}
%     \centering
%     \includegraphics[height=42mm, width=80mm]{Figures/SR_HO_rho.eps}
%     \vspace{-2mm}
% \captionsetup{font=small}
%     \caption{Sum-rate and CC number vs multi-objective factor $\rho$.}
%     \label{fig:SR_HO_rho}
% \end{figure}

\begin{figure}
    \centering
    \includegraphics[height=58mm, width=85mm]{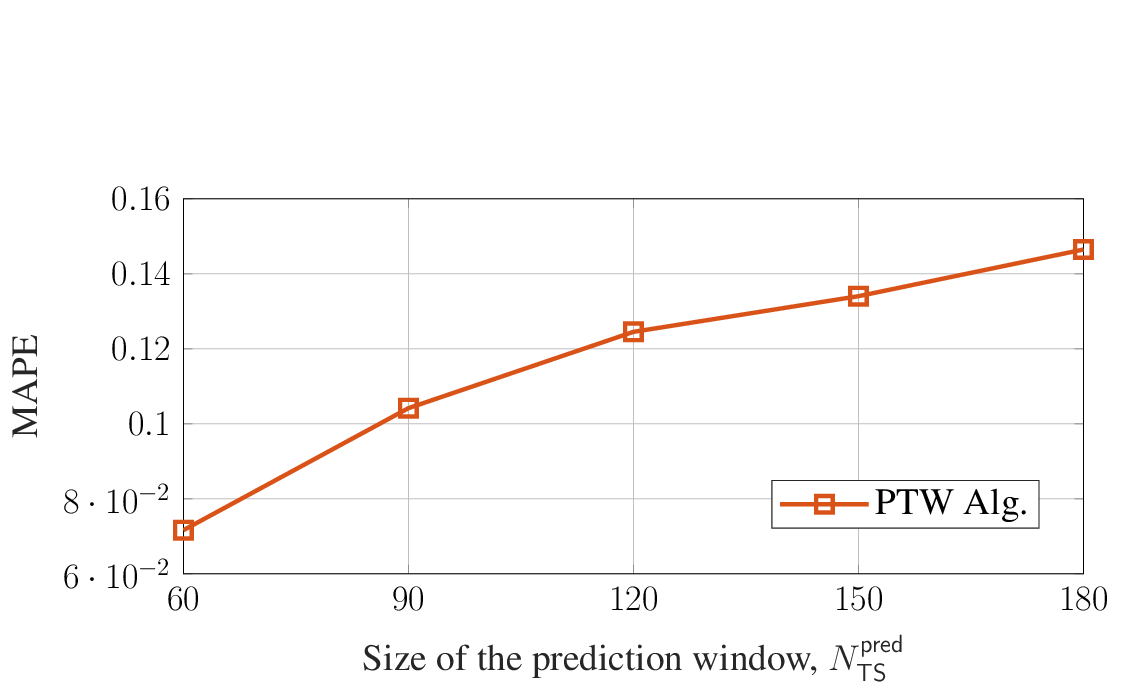}
    \vspace{-2mm}
\captionsetup{font=small}
    \caption{Mean absolute percentage error (MAPE) of Alg.~\ref{alg: pred chan} vs. prediction sub-TW size.}
    \label{fig:MAPE_Ntsp}
\end{figure}

\begin{figure}
    \centering
    \includegraphics[width=90mm]{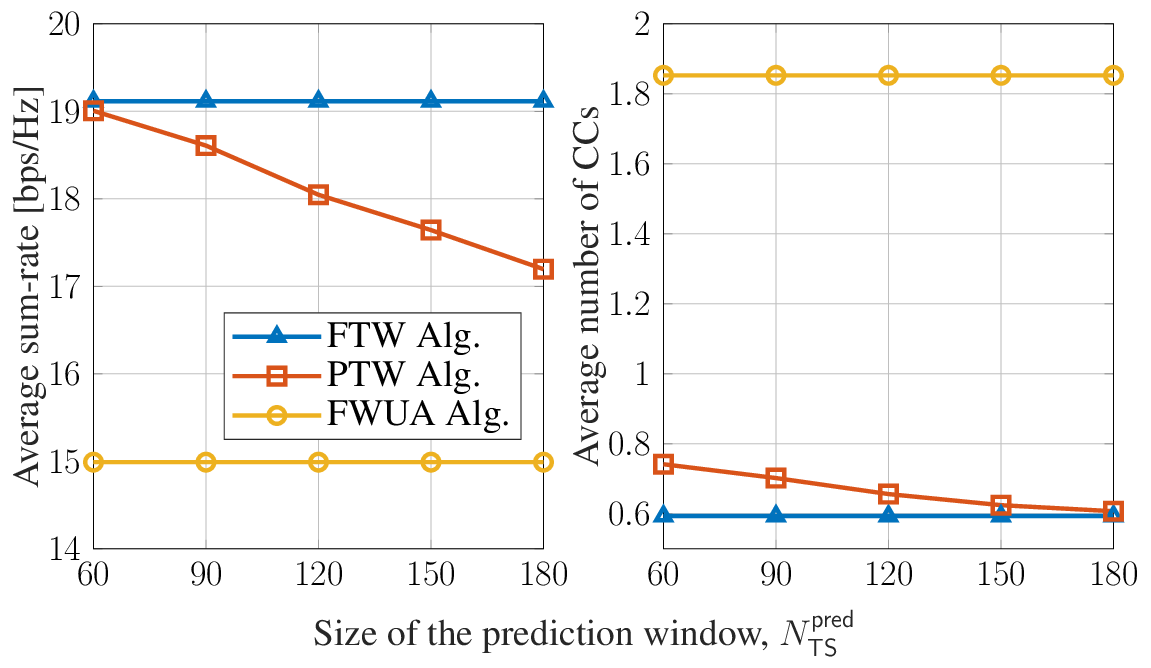}
    \vspace{-2mm}
\captionsetup{font=small}
    \caption{Sum-rate and CC number of Alg.~\ref{alg: OpzPower Pred} vs. prediction sub-TW size.}
    \label{fig:SR_HO_Ntsp}
\end{figure}

Fig.~\ref{fig:MAPE_Ntsp} shows the statistic channel prediction error in terms of the mean absolute percentage error (MAPE).
Obviously, a larger prediction window leads to a higher channel prediction error. Particularly, the MAPE values are about $0.07$ and $0.14$ at $N_{\sf{TS}}^{\sf{pred}} = 60$ TSs and $N_{\sf{TS}}^{\sf{pred}} = 180$ TSs, respectively.
For more information on the impact of prediction window size, Fig.~\ref{fig:SR_HO_Ntsp} illustrates the performance of the PTW Algorithm in terms of the average SR and the CC number with the different sizes of prediction sub-TW $N_{\sf{TS}}^{\sf{pred}}$ and those of the FTW and FWUA algorithms as the baselines. Regarding the CC, since the sub-TW is optimized independently, a longer prediction window size leads to a lower number of sub-TW, thus the CC number decreases. However, a larger window size results in less accuracy in channel gain prediction as MAPE increases, leading to degradation of SR performance. Additionally, the UA decision is made based on the predicted channel, the higher prediction error may lead to more unsuitable UA decisions with the actual channel, which also contributes to the SR degradation. 
For instance, the SR degradation is about $2$ bps/Hz and the average number of CCs decreases about $0.15$ when raising $N_{\sf{TS}}^{\sf{pred}}$ from $60$ to $180$. This can be considered as the trade-off between SR and CC performances. However, these performances can be balanced and comparable with those of the FTW Algorithm by choosing an appropriate prediction sub-TW size.

\begin{figure}
    \centering
    \includegraphics[width=90mm]{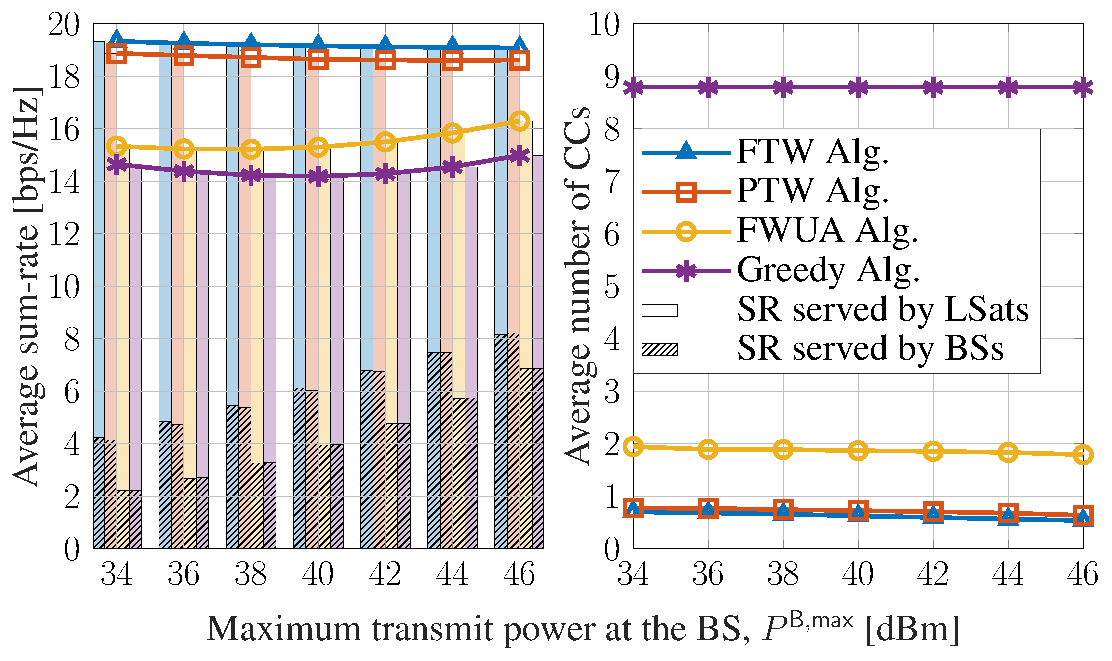}
    \vspace{-2mm}
\captionsetup{font=small}
    \caption{Sum-rate and CC number vs. BS maximum transmit power.}
    \label{fig:SR_HO_PBS}
\end{figure}

Fig.~\ref{fig:SR_HO_PBS} depicts the impact of the BS's power budget on the average SR and number of CCs. Regarding the CC results, one can see that the change in BS's power budget does not significantly affect the average number of CCs. However, two proposed FTW and PTW Algorithms outperform the two other benchmarks in terms of average CC number. In particular, the average CC number outcome of the two proposed algorithms is about $0.65$ whereas those of the FWUA and Greedy Algorithms are about $1.9$ and $8.7$, respectively. Regarding the SR, the trends of the FWUA and Greedy Algorithms first decrease and then increase as the BS's power budget increases. This phenomenon arises because interference from BSs to LSat-UE links increases with the BS's power budget, becoming significant at particular BS power levels. Subsequently, at a sufficiently high BS power, the SR offered by BS-UE links dominates that by LSat-UE links, which leads to an increase in system SR. 
Additionally, increasing the BS's power budget can improve the BS-UE links. This is shown by the increase in the ratio of the SR contributed by BSs with the BS's power budget. Regarding the two proposed algorithms, the ratio is about $22$\% at $P^{\sf{B,max}}=34$ dBm and $43$\% at $P^{\sf{B,max}}=46$ dBm.
Besides, due to the flexibility of PA in the FTW and PTW algorithms, the two proposed algorithms offer a significant SR improvement. Especially, the SR gap between the two lines is not significant, i.e., about $0.5$ bps/Hz. For instance, SR outcomes of the FTW and PTW algorithms are about $19.1$ and $18.6$ bps/Hz at $P^{\sf{B,max}}=40$ dBm, respectively.

\begin{figure}
    \centering
    \includegraphics[width=90mm]{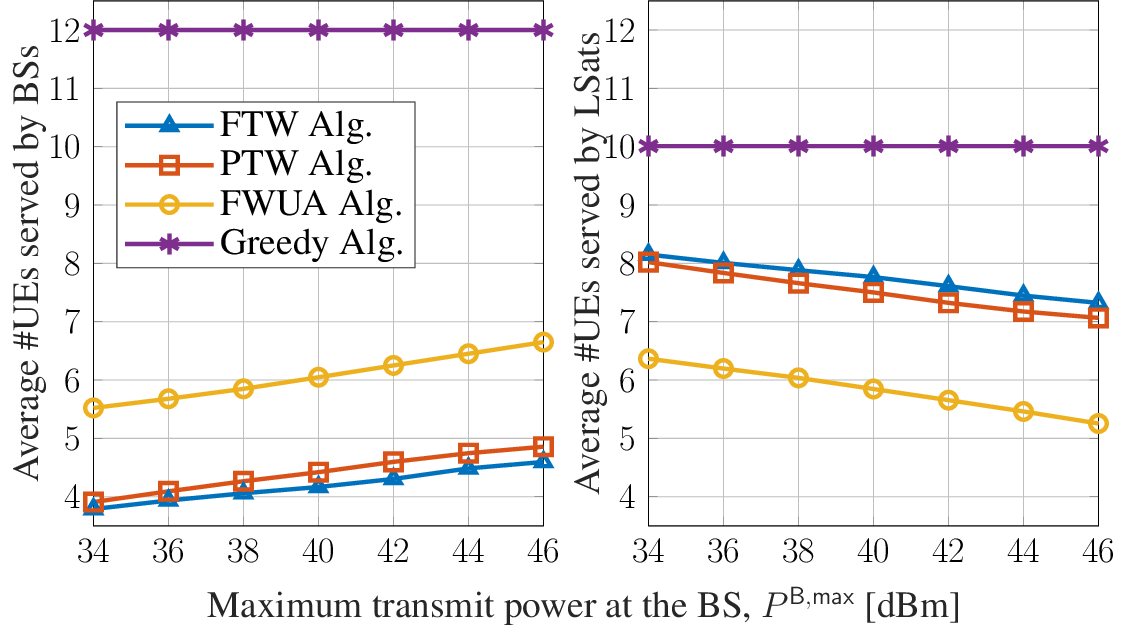}
    \vspace{-2mm}
\captionsetup{font=small}
    \caption{Served-UE number vs BS maximum transmit power.}
    \label{fig:NumConnec_PBS}
\end{figure}

For more insight into the impact of BS's power budget, Fig.~\ref{fig:NumConnec_PBS} displays the average number of UEs served by BSs and LSats with the different $P^{\sf{B,max}}$. Regarding the two proposed algorithms, one can see that the average number of UEs served in BSs increases as $P^{\sf{B,max}}$ increases. For instance, that is about $3.8$ and $4.7$ UEs at $P^{\sf{B,max}} = 34$ dBm and $P^{\sf{B,max}}=46$ dBm, respectively.
Additionally, that number at LSats decreases as $P^{\sf{BS,max}}$ increases, which is about $8.1$ and $7.3$ UEs at $P^{\sf{B,max}} = 34$ dBm and $P^{\sf{B,max}}=46$ dBm, respectively. 
This indicates that UEs priories BSs's connections to achieve a higher SR as the BS's power budget increases. 
Furthermore, the average numbers of served UEs at BSs and LSats provided by the proposed FTW and PTW Algorithms are less than that by the greedy algorithm. This emphasizes the effectiveness of the proposed algorithms wherein they require less number of connections while providing better SR compared to the two benchmark ones, i.e., the FWUA and greedy algorithms.

\begin{figure}
    \centering
    \includegraphics[width=90mm]{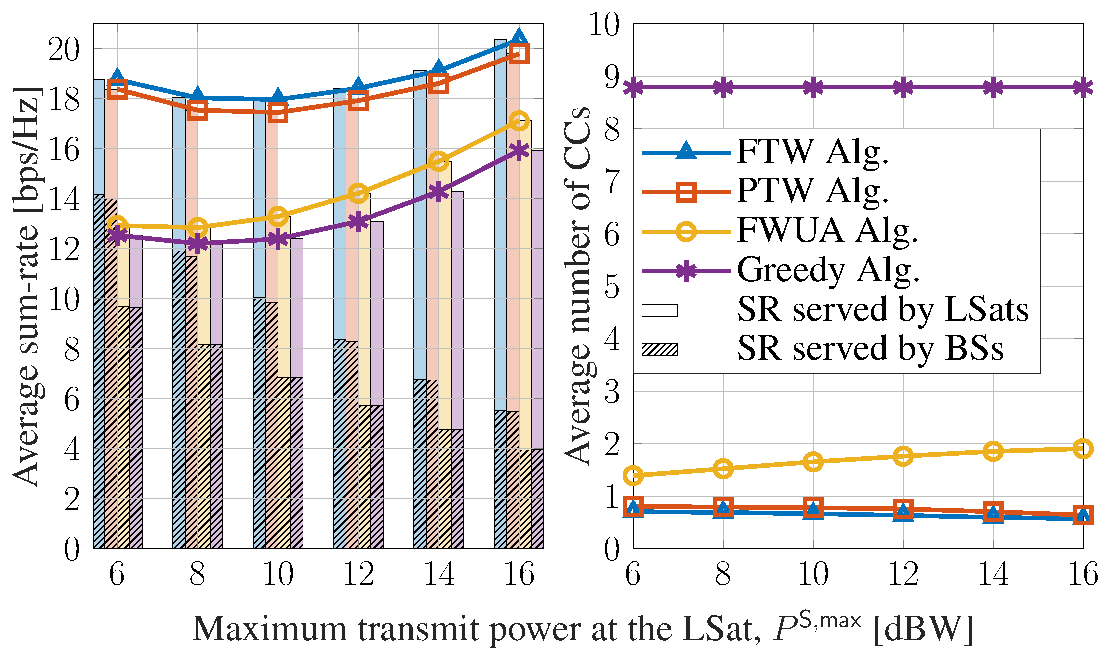}
    \vspace{-2mm}
\captionsetup{font=small}
    \caption{Sum-rate and the number of CCs versus the maximum transmit power at the LSat.}
    \label{fig:SR_HO_PSat}
\end{figure}

Fig.~\ref{fig:SR_HO_PSat} shows the average SR and number of CCs versus the LSat's power budget. Regarding the CC number, this is similar to that in Fig.~\ref{fig:SR_HO_PBS}, wherein its change is not significant as different $P^{\sf{S,max}}$. Regarding the SR performance, the trends of the four considered algorithms are similar. In particular, as the power budget at LSats increases, the SR first drops due to the increase of interference from LSats to BS-UE links. Subsequently, the SR rises at a sufficiently high LSat power budget due to the domination quality of LSat-UE links compared to that of BS-UE links. For instance, at $P^{\sf{S,max}}=(6, 12, 16)$ dBW, the SR provided by the FWT and PTW Algorithms is approximately $(18.8, 18.4, 20.4)$ bps/Hz and $(18.4, 17.9, 19.8)$ bps/Hz, respectively. 
Furthermore, the SR contribution ratio of LSats increases with the LSat's power budget due to the LSat-UE link's improvement. Particularly, regarding the two proposed algorithms, LSats contributes about $24$\% and $74$\% of SR at $P^{\sf{S,max}}=6$ dBW and $P^{\sf{S,max}}=16$ dBW, respectively.
Compared to SR at different $P^{\sf{BS,max}}$ in Fig.~\ref{fig:SR_HO_PBS}, the SR degradation in this figure shows a more critical impact of interference from LSats to BS-UE links, even with PA of the two proposed algorithms. This is explained by the impact of interference from LSats on BS-UE links in a vast area, whereas BSs cause interference on LSat-UE links locally due to the isolation by high buildings.
% \vspace{-1mm}

\begin{figure}
    \centering
    \includegraphics[width=90mm]{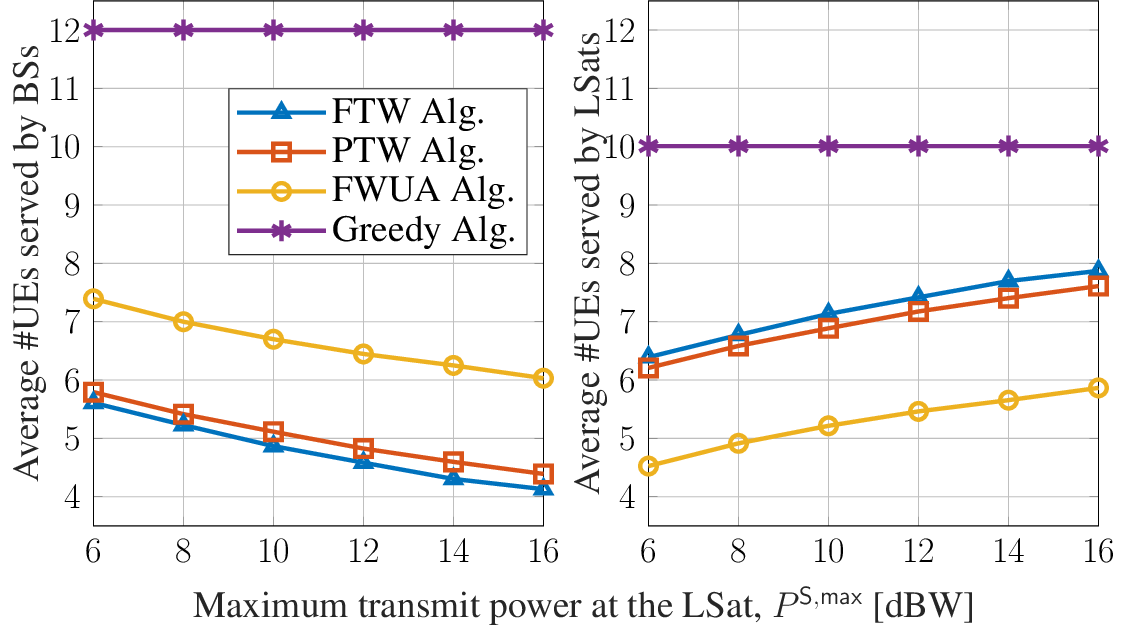}
      \vspace{-2mm}
\captionsetup{font=small}
    \caption{Average served-UE number vs LSat max transmit power.}
    \label{fig:NumConnec_PSat}
    \vspace{-3mm}
\end{figure}

For more insight into the impact of LSat's power budget, Fig.~\ref{fig:NumConnec_PSat} shows the average number of served UEs at BSs and at LSats with the different cases of $P^{\sf{S,max}}$. In the two proposed algorithms, the UEs reduce the connections to the BSs and increase those at LSats as the maximum LSat power increases since the SR offered by LSats is improved. Particularly, the average numbers of UEs served at BSs and LSats are about $5.7$ and $6.3$ at $P^{\sf{S,max}} = 6$ dBW, and about $4.2$ and $7.8$ at $P^{\sf{S,max}} = 16$ dBW. Furthermore, compared to the benchmark algorithms, this figure again shows the effectiveness of the two proposed algorithms in terms of SR and connection number wherein their solutions have similar or fewer numbers of connections but better SR performance.

\begin{figure}
    \centering
    \includegraphics[width=90mm]{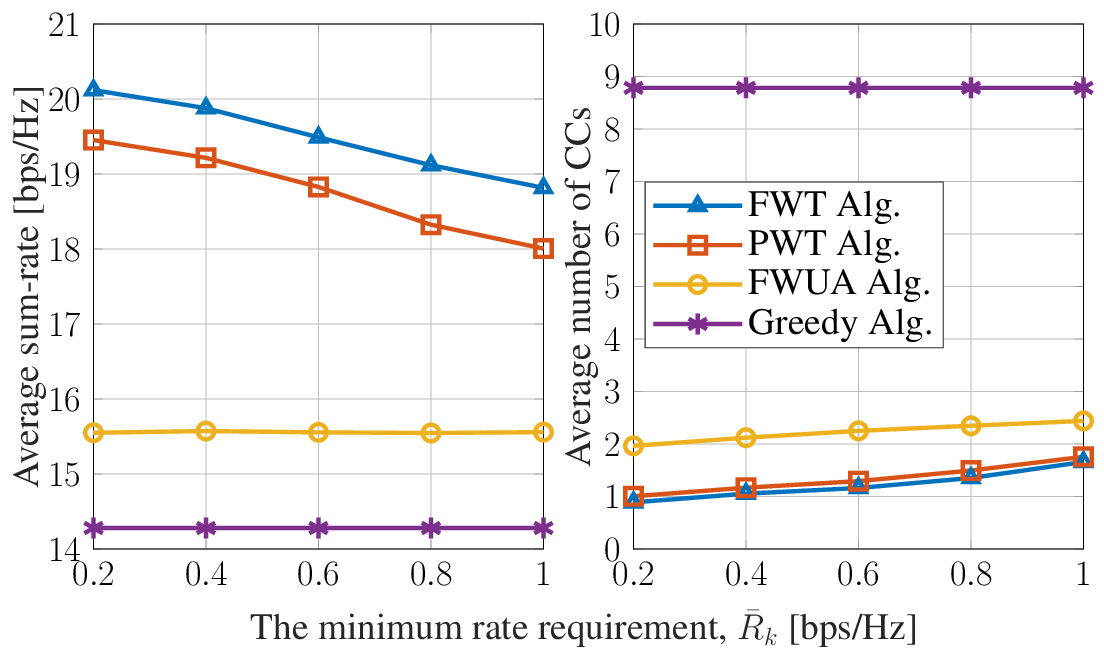}
    \vspace{-2mm}
\captionsetup{font=small}
    \caption{Sum-rate and CC number versus the minimum rate requirement.}
    \label{fig:SR_HO_Rmin}
\end{figure}

Subsequently, Fig.~\ref{fig:SR_HO_Rmin} illustrates the average SR and the number of CCs versus the different rate requirement $\bar{R}_{k}$.
Regarding the FWUA algorithm, the SR performance changes insignificantly with different rate requirements. However, to satisfy the QoS constraint, the number of CCs increases as $\bar{R}_{k}$ increases.
For the two proposed algorithms, as $\bar{R}_{k}$ increases, the SR outcomes decrease whereas the numbers of CCs increase. One can see that the increase in the rate requirement leads to a smaller feasible set of the SR-CC problem, which results in degradation in the SR and CC performance. Particularly, at $\bar{R}_{k}=0.2$ bps/Hz, the average SR results of the FTW and PTW algorithms are about $20.12$ bps/Hz and $19.45$ bps/Hz, and the CC number is about $0.9$; at $\bar{R}_{k}=1$ bps/Hz, the those SR results are about $18.8$ bps/Hz and $18.1$ bps/Hz, and the CC number is about $1.7$, respectively. 
However, at all considered rate requirement levels, the two proposed algorithms always outperform the FWUA and greedy algorithms in terms of SR and the CC-number.

\begin{figure}
    \centering
    \includegraphics[width=90mm]{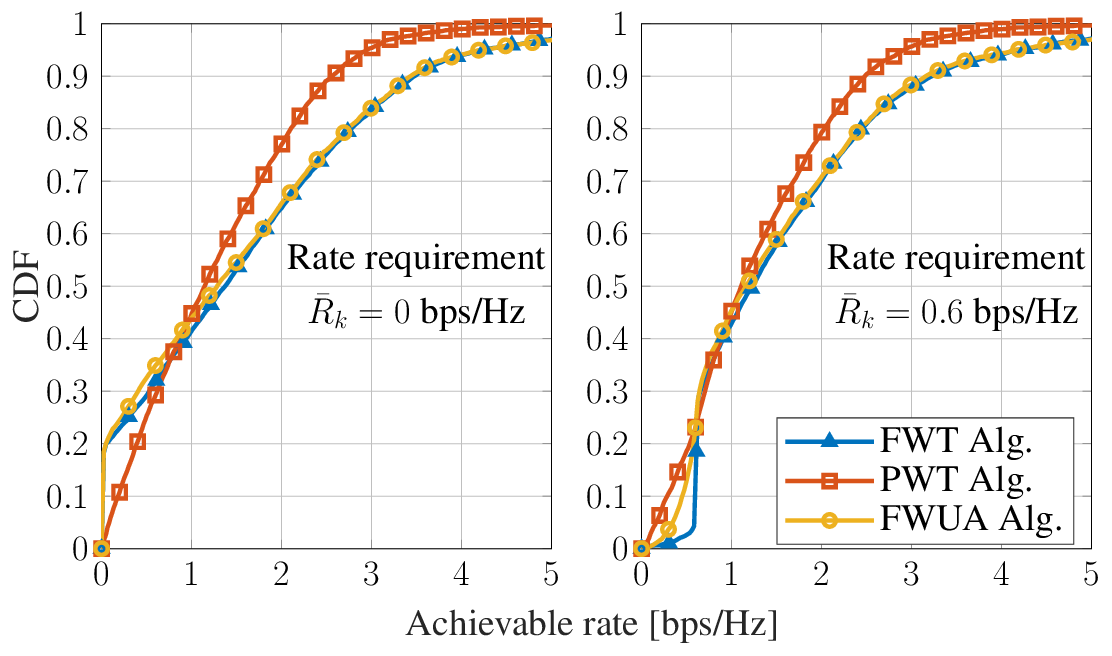}
    \vspace{-2mm}
\captionsetup{font=small}
    \caption{The cumulative distribution function of the achievable rate at the minimum rate requirements of $\bar{R}_{k}=0$ bps/Hz and $\bar{R}_{k}=0.6$ bps/Hz.}
    \label{fig:CDF_Rmin}
\end{figure}

For more information on the impact of QoS requirements, Fig.~\ref{fig:CDF_Rmin} shows the cumulative distribution function (CDF) of the achievable rate with rate requirement $\bar{R}_{k}=0$ bps/Hz and $\bar{R}_{k}=0.6$ bps/Hz. Regarding the FWUA algorithm, the CDF of achievable rates at  $\bar{R}_{k}=0$ bps/Hz and $\bar{R}_{k}=0.6$ bps/Hz are similar, which further clarifies the near constancy of SR in Fig.~\ref{fig:SR_HO_Rmin}. In contrast, the CDFs of the rate outcomes of the FTW and PTW algorithms change significantly at $\bar{R}_{k}=0$ bps/Hz and $\bar{R}_{k}=0.6$ bps/Hz. Particularly, compared to the CDFs at $\bar{R}_{k}=0$ bps/Hz, those at $\bar{R}_{k}=0.6$ bps/Hz are forced to satisfy the minimum rate requirement. However, due to the channel prediction error in the PWT algorithm, one can see the degradation of the achievable rate, especially at lower than $\bar{R}_{k}=0.6$ bps/Hz. However, the probabilities of rate outcomes of the FTW and PTW algorithms lower than $\bar{R}_{k}=0.6$ bps/Hz are relatively small, i.e., about $0.03$ and $0.2$, respectively.

\begin{figure}
    \captionsetup{font=small}
    \centering
    \includegraphics[width=80mm]{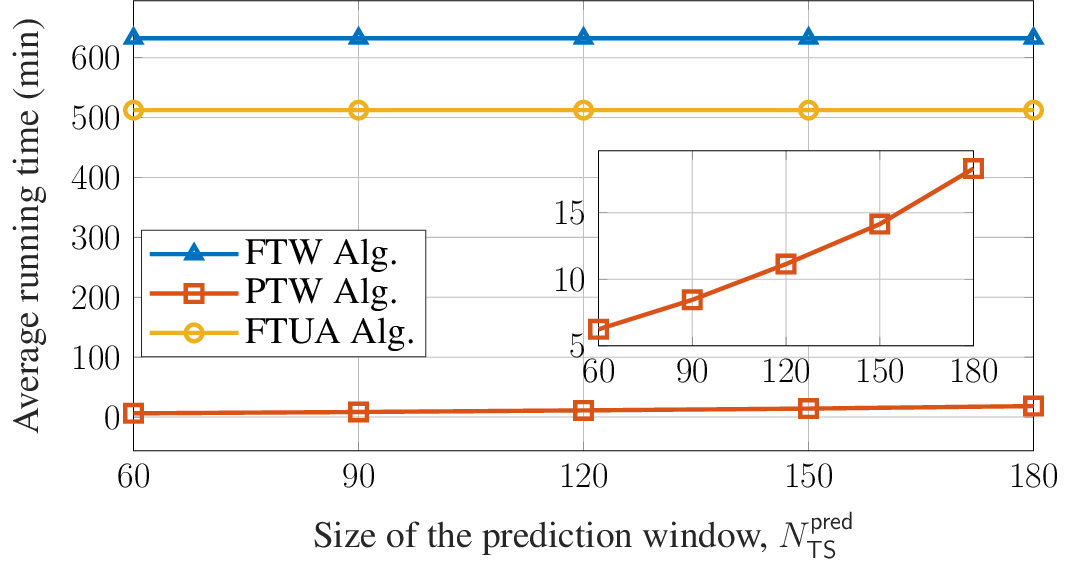}
    \vspace{-2mm}
    \caption{Average running time of examined iterative algorithms.}
    \label{fig:time_Ntsp}
\end{figure}

To assess computational efficiency, the average running time of the PTW algorithm with varying prediction window sizes $N_{\sf{TS}}^{\sf{pred}}$ and those of the benchmark, i.e., the FTW and FTUA algorithms, are shown in Fig.~\ref{fig:time_Ntsp}. Since the FTW and FTUA algorithms solve the problem over the entire time window, they incur an extremely high time for execution, i.e., about $632.3$ and $512.4$ minutes.
In contrast, the PTW algorithm exhibits much lower running time. While its running time increases slightly with larger prediction window size, due to the increasing size of problem $\mathcal{(P_{2})_{\kappa}}$, it remains substantially more efficient than the FTW and FTUA ones. Specifically, for $N_{\sf{TS}}^{\sf{pred}} = (60,90,120,150,180)$~TSs, the average running time of the PTW algorithm is approximately $6.2$, $8.4$, $11.1$, $14.2$, and $18.31$ minutes, respectively. Notably, the proposed PTW algorithm can achieve performance comparable to the FTW algorithm and even outperforms the FTUA one in terms of max-SR and min-CC-number, while requiring only a fraction of the running time, i.e, approximately $1\% - 3.5 \%$ of the benchmarks' running time.

%\vspace{-3mm}
\subsection{Practical Implementation Analysis}
\begin{figure}
    \centering
    \includegraphics[width=85mm]{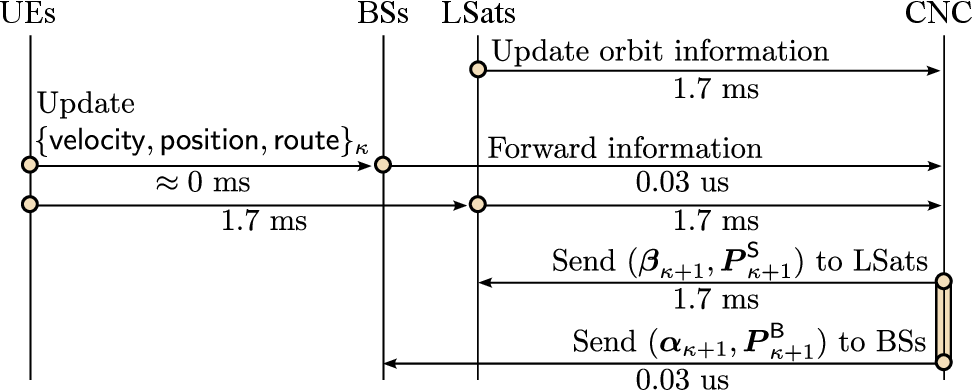}
    \vspace{-2mm}
    \captionsetup{font=small}
    \caption{Propagation and signaling time consumption.}
    \label{fig:PredAlg_time}
\end{figure}

According to the simulation setting, assuming that the CNC is deployed at the center of the examined area, the distance between the CNC and BSs is about $7.8$ km leading to the propagation time between CNC and BSs of $T_{\sf{CNC,BS}} \leq 0.03$~\textmu s. Besides, due to the satellite orbit altitude of $500$ km, the propagation time between the CNC and LSats is about $T_{\sf{CNC,LSat}} \approx 1.7$ ms. The time consumption due to propagation and signaling is depicted in Fig.~\ref{fig:PredAlg_time}. In the best case where UEs can update their status via BSs, this time consumption is about $T_{\sf{best}} = 2 \times T_{\sf{CNC,LSat}} = 3.4$ ms, while in the worst case where existing UEs which need to update their status via LSats this time is about $T_{\sf{worst}} = 3 \times T_{\sf{CNC,LSat}} = 5.1$ ms. Additionally, for TS duration of $T_{\sf{S}} = 0.5$ s and the sub-TW size of $N_{\sf{TS}}^{\sf{pred}} = 90$ TSs as in simulation setting, the sub-TW duration is $T_{\sf{sub-TW}} = 45$ s. One can see that the time consumption for propagation and signaling in the worst case is negligible compared to sub-TW duration. Hence, this further shows the practicality of the proposed PTW algorithm.

Besides, the segmentation of the map in the considered area can reduce the complexity of the prediction mechanism which facilitates scalability. Particularly, based on the outcome of the UE position prediction step, for each UE, one can determine the map segment corresponding to its position in a sub-TW. Based on which the RT mechanism can execute for the identified map segment rather than the entire considered map. Furthermore, these steps can be executed in a parallel manner for UEs which further facilitates scalability.

%\vspace{-4mm}
\section{Conclusions} \label{sec: concl}
\vspace{-1mm}
This work studied the joint UA and PA for the practical co-primary spectrum sharing ISTNs in urban environments. Subsequently, the multi-objective of SR maximization and the number of CCs minimization under the resource limitation constraints is formulated. To solve this problem, we proposed two algorithms: 1) FTW: an iterative algorithm for the entire time-window and 2) PTW: a prediction-based algorithm for the implementation purpose that sequentially predicts the channel gain and solves the problem for sub-time-windows. For practice evaluation, simulation with the actual 3D map and UE mobility is performed. The simulation results provide insight into the link budget assessment in urban environments. Additionally, numerical results show the effectiveness of our two proposed algorithms compared to the greedy algorithm and the benchmark one (FTUA) in terms of both the system SR and the CC number.

The approach of using RT, 3D map, and realistic mobility can be used as a framework for future works. Moreover, this study provides opportunities for further extensions, particularly in improving the channel model as well as addressing prediction error and intra-system-interference issues. Specifically, for practical applications, random impacts in the actual environment can be characterised in the channel model by incorporating corresponding statistical components. Besides, the prediction mechanism in the proposed PTW algorithm could be improved to reduce prediction error and SR degradation. For more in-depth analysis, the potential intra-system-interference issues should be considered in future research.
% This work studied the joint UA and PA for the practical co-primary spectrum sharing ISTNs in urban environments. Subsequently, the multi-objective of SR maximization and number of HOs minimization under the resource limitation constraints is formulated. To solve this problem, we proposed an iterative algorithm for the entire time-window. Furthermore, for the implementation purpose, we proposed a prediction-based algorithm to sequentially predict channel gain and solve the problem for sub-time-windows. The simulation with the actual 3D map and UE mobility are conducted, which shows the effectiveness of our proposed algorithms compared to the greedy and benchmark algorithms in terms of both SR and CC number performances.

% \section*{Acknowledgment}
% \vspace{-1mm}
% This work has been supported by the Luxembourg National Research Fund (FNR) under the project INSTRUCT (IPBG19/14016225/INSTRUCT). 
% %and project MegaLEO (C20/IS/14767486).

\appendices
\vspace{-4mm}
\section{Proof of Proposition~\ref{pro: up bound fCC}} \label{app: up bound fCC}
\vspace{-1mm}
Consider function $f_{\sf{abs}}(x,y) = |\Vert x \Vert_{0} - \Vert y \Vert_{0}|, \; x,y \geq 0$ which has the form of components in the CC-number function. $f_{\sf{abs}}(x,y)$ can be represented as $f_{\sf{abs}}(x,y) = \max\{\Vert x \Vert_{0},\Vert y \Vert_{0}\} - \min\{\Vert x \Vert_{0},\Vert y \Vert_{0}\}$. Let introduce slack variables $u^{\sf{low}}$ and $u^{\sf{up}}$ satisfying
\bieq{ll} \label{eq: abs apx}
    u^{\sf{low}} \leq \Vert x \Vert_{0}, \; u^{\sf{low}} \leq \Vert y \Vert_{0}, \subnum \label{eq: abs apx a} \\
    u^{\sf{up}} \geq \Vert x \Vert_{0}, \; u^{\sf{up}} \geq \Vert y \Vert_{0}, \subnum \label{eq: abs apx b}
\eieq
an upper bound of function $f_{\sf{abs}}(x,y)$ can be obtained as
\beq
    f_{\sf{abs}}(x,y) \leq u^{\sf{up}} - u^{\sf{low}}.
\eeq
Additionally, thanks to the approximations in \eqref{eq: norm0 apx} and \eqref{eq: up bound fapx}, constraint \eqref{eq: abs apx} is approximated as
\bieq{ll} \label{eq: abs apx 2}
    u^{\sf{low}} \leq f_{\sf{apx}}(x), \; u^{\sf{low}} \leq f_{\sf{apx}}(y), \subnum \label{eq: abs apx 2a} \\
    u^{\sf{up}} \geq f_{\sf{apx}}^{(i)}(x), \; u^{\sf{up}} \geq f_{\sf{apx}}^{(i)}(y). \subnum \label{eq: abs apx 2b}
\eieq
Constraint \eqref{eq: abs apx 2} is convex since $f_{\sf{apx}}(\cdot)$ is concave and $f_{\sf{apx}}^{(i)}(\cdot)$ is affine. Subsequently, substituting this approximation for each absolute term in the CC-number function $\bar{f}^{\sf{CC}}(\boldsymbol{P}^{\sf{B}}, \boldsymbol{P}^{\sf{S}})$ setting $x = P_{n,k,t}^{\sf{B}}$, $y = P_{n,k,t-1}^{\sf{B}}$, $u^{\sf{low}} = a_{n,k,t}^{\sf{low}}$, and $u^{\sf{up}} = a_{n,k,t}^{\sf{up}}$ for the term $| \Vert P_{n,k,t}^{\sf{B}} \Vert_{0} - \Vert P_{n,k,t-1}^{\sf{B}} \Vert_{0} |$, and $x = P_{m,k,t}^{\sf{S}}$, $y = P_{m,k,t-1}^{\sf{S}}$, $u^{\sf{low}} = b_{m,k,t}^{\sf{low}}$, and $u^{\sf{up}} = b_{m,k,t}^{\sf{up}}$ for term $| \Vert P_{m,k,t}^{\sf{S}} \Vert_{0} - \Vert P_{m,k,t-1}^{\sf{S}} \Vert_{0} |$, we obtain the upper bound $\tilde{f}^{\sf{CC},(i)}(\boldsymbol{a}, \boldsymbol{b})$ of the CC-number function $\bar{f}^{\sf{CC}}(\boldsymbol{P}^{\sf{B}}, \boldsymbol{P}^{\sf{S}})$ as in \eqref{eq: fCC 3} and convex constraints $(\tilde{C}11),(\tilde{C}12)$.

\vspace{-3mm}
\section{Proof of Proposition~\ref{pro: convexify rate}} \label{app: convexify rate}
{ Let consider function $f_{\sf{rate}}(x,y) = \ln(1+\frac{x}{y + c})$, $x,y \geq 0, c > 0$, and constraint $f_{\sf{rate}}(x,y) \geq z$. By expressing $f_{\sf{rate}}(x,y) = \ln(x + y + c) - \ln(y + c)$, constraint $f_{\sf{rate}}(x,y) \geq z$ can be decomposed as}
\begin{subnumcases}{}
    \ln(x + y + c) \geq z + u, \label{eq: apx rate 1a} \\
    \ln(y + c) \leq u, \label{eq: apx rate 1b}
\end{subnumcases}
wherein $z$ is a slack variable. By taking the exponent operator on both sides of \eqref{eq: apx rate 1b}, it is transformed into
\beq \label{eq: apx rate 2b}
y + c \leq \exp(u). 
\eeq
It can be seen that \eqref{eq: apx rate 2b} is still non-convex. To convexify \eqref{eq: apx rate 2b}, we employ the first-order approximation to the exponential term and obtain the following convex constraint
\begin{subnumcases}{}
    \ln(x + y + c) \geq z + u , \label{eq: apx rate 3a} \\
    y + c \leq \exp(u^{(i)})(u - u^{(i)} + 1), \label{eq: apx rate 3b}
\end{subnumcases}
wherein $u^{(i)}$ is the feasible point of $u$ at iteration $i$.

By setting
% \bieq{ll}
$x = P_{n,k,t}^{\sf{B}} h_{n,k,t}$,
$y = \sum_{\forall m} \sum_{\forall k'} P_{m,k',t}^{\sf{S}} g_{m,k,t}$, 
$c = \sum_{\forall m} \eta_{m,t}^{\sf{S}} \bar{P}_{m,t}^{\sf{S}} g_{m,k,t} + \sigma_{k}^2$, 
$z = \lambda_{n,k,t}^{\sf{B}}$, and
$u = \mu_{k,t}^{\sf{B}}$,
% \eieq
we obtained the convex approximated form of constraint $(C8)$ as constraints $(\tilde{C}8.1)-(\tilde{C}8.2)$.
Subsequently, by setting
% \bieq{ll}
$x = P_{m,k,t}^{\sf{S}} g_{m,k,t}$, 
$y = \sum_{\forall n} \sum_{\forall k'} P_{n,k',t}^{\sf{B}} h_{n,k,t}$,
$c = \sum_{\forall n} \eta_{n,t}^{\sf{B}}  \bar{P}_{n,t}^{\sf{B}} h_{n,k,t} + \sigma_{k}^2$,
$z = \lambda_{m,k,t}^{\sf{S}}$, and
$u = \mu_{k,t}^{\sf{S}}$,
% \eieq
the convex approximated form of constraint $(C9)$ is derived as $(\tilde{C}9.1) - (\tilde{C}9.2)$.

\vspace{-1mm}
\bibliographystyle{IEEEtran}
%\balance
\bibliography{Journal}
\end{document}